\documentclass[sigconf]{acmart}
\AtBeginDocument{%
  }

\usepackage{array} % Add this in your preamble
\usepackage{subcaption} % add this in your preamble
\usepackage{xcolor}

\definecolor{brightboldred}{RGB}{255, 20, 20}

\newif\ifshowchanges
\showchangesfalse     % <- CLEAN VERSION (no red)

\ifshowchanges
    \newcommand{\red}[1]{\textbf{\textcolor{brightboldred}{#1}}}
\else
    \newcommand{\red}[1]{#1}
\fi
\copyrightyear{2026}
\acmYear{2026}
\setcopyright{cc}
\setcctype{by}
\acmConference[CHI '26]{Proceedings of the 2026 CHI Conference on Human Factors in Computing Systems}{April 13--17, 2026}{Barcelona, Spain}
\acmBooktitle{Proceedings of the 2026 CHI Conference on Human Factors in Computing Systems (CHI '26), April 13--17, 2026, Barcelona, Spain}
\acmPrice{}
\acmDOI{10.1145/3772318.3791439}
\acmISBN{979-8-4007-2278-3/2026/04}

\begin{document}

\title{How Does Delegation in Social Interaction Evolve Over Time? Navigation with a Robot for Blind People}

\author{Rayna Hata}
\email{rhata@andrew.cmu.edu}
\affiliation{%
  \institution{Carnegie Mellon University}
  \city{Pittsburgh}
  \state{Pennsylvania}
  \country{USA}
}

\author{Masaki Kuribayashi}
\email{rugbykuribayashi@toki.waseda.jp}
\affiliation{%
  \institution{Waseda University}
  \city{Tokyo}
  \country{Japan}
}
\affiliation{%
  \institution{Miraikan - The National Museum of Emerging Science and Innovation}
  \city{Tokyo}
  \country{Japan}
}

\author{Allan Wang}
\email{allan.wang@jst.go.jp}
\affiliation{%
  \institution{Miraikan - The National Museum of Emerging Science and Innovation}
  \city{Tokyo}
  \country{Japan}
}

\author{Hironobu Takagi}
\email{takagih@jp.ibm.com}
\affiliation{%
  \institution{IBM Research - Tokyo}
  \city{Tokyo}
  \country{Japan}
}

\author{Chieko Asakawa}
\email{chiekoa@us.ibm.com}
\affiliation{%
  \institution{Miraikan - The National Museum of Emerging Science and Innovation}
  \city{Tokyo}
  \country{Japan}
}

\newcommand{\boldparagraph}[1]{\vspace{0.2cm}\noindent{\bf #1:}}
\newcommand{\todo}[1]{\textcolor{red}{\textbf{[#1]}}}
\begin{abstract}
Autonomy and independent navigation are vital to daily life but remain challenging for individuals with blindness. Robotic systems can enhance mobility and confidence by providing intelligent navigation assistance. However, fully autonomous systems may reduce users’ sense of control, even when they wish to remain actively involved. Although collaboration between user and robot has been recognized as important, little is known about how perceptions of this relationship change with \red{repeated} use. We present a \red{repeated exposure study} with six blind participants who interacted with a navigation-assistive robot in a real-world museum. Participants completed tasks such as navigating crowds, approaching lines, and encountering obstacles. Findings show that participants refined their strategies over time, developing clearer preferences about when to rely on the robot versus act independently. This work provides insights into how strategies and preferences evolve with repeated interaction and offers design implications for robots that adapt to user needs over time.
\end{abstract}






\begin{teaserfigure}
  \includegraphics[width=\textwidth, trim=0mm 0mm 5mm 0mm, clip]{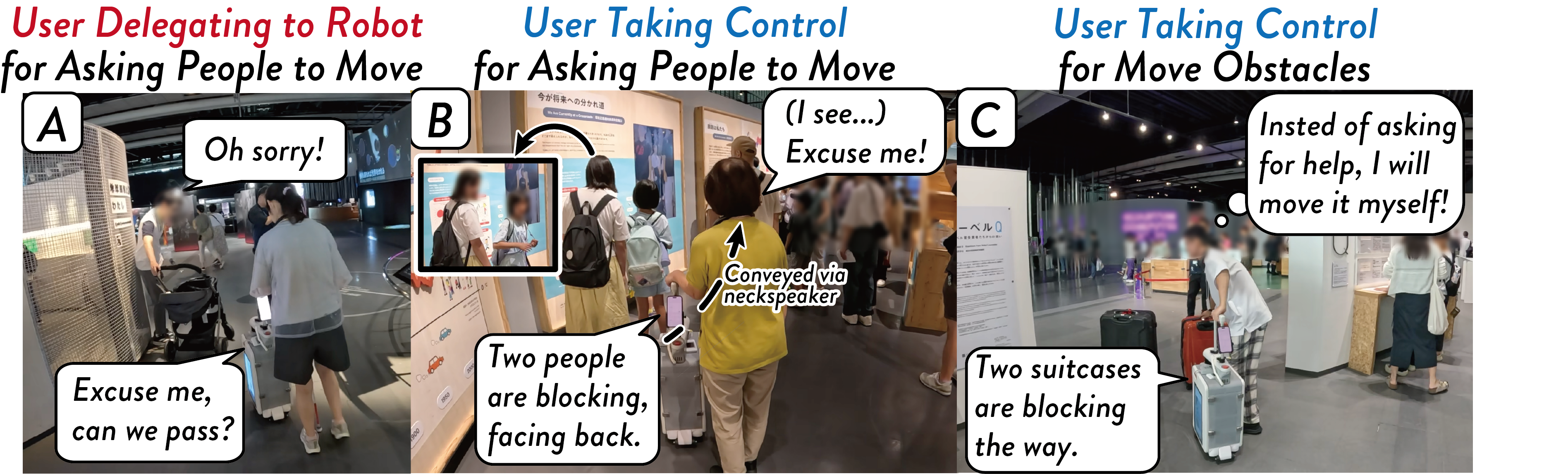}
  \caption{\textbf{User Delegation vs. User Taking Control.} Examples of how users managed social interactions during navigation. In (A), the user delegates to the robot, which asks people with a stroller to move. In (B), the user takes control by speaking and asking the two people to move. In (C), the user again takes control, deciding to move the blocking suitcases without asking for help.}
  \Description{Three-panel photo sequence comparing how users handle blocked paths when navigating with a robot. Panel A shows “User Delegating to Robot,” where the robot asks people with a stroller to move and they step aside. Panel B shows “User Taking Control,” where the robot detects two people blocking the way and the user uses a neck speaker to say “Excuse me.” Panel C shows another “User Taking Control” case, where suitcases block the way and the user decides to move them instead of asking for help.}
  \label{fig:teaser}
\end{teaserfigure}

\maketitle

\section{Introduction}

Autonomy and independent navigation are vital components of daily life, yet they remain challenging for individuals with blindness. Robotic systems have been explored to enhance mobility and confidence by providing autonomous navigation assistance in complex environments~\cite{bonani2018my,doore2024co,wachaja2017navigating,kayukawa2020blindpilot}. 

Nonetheless, there are situations where the robot cannot navigate successfully on its own, such as when encountering dense crowds or physical obstacles~\cite{wang2023social}. Collaboration between the user and the robot is essential for overcoming such challenging scenarios. Shared control, an emerging interaction paradigm in which both user input and robot autonomy jointly shape navigation or task completion decisions, provides a way to address situations that autonomous systems alone cannot resolve~\cite{music2017control,amirshirzad2019human, wang2023social}.  

In human-computer interaction (HCI) and human-robot interaction (HRI), shared control has been widely studied as a way to balance efficiency with user agency. For blind users, robot autonomy is often prioritized for convenience and safety; however, prior findings suggest that many also value social autonomy and prefer to remain actively engaged in decision-making and interaction with the surroundings, choosing when and how to rely on robotic support~\cite{kamikubo2025beyond}.

On the spectrum between user-led and robot-led interaction in navigation-assistive contexts, \textit{delegation} leans toward the robot-led side. For example, when blocked by a crowd, delegation involves the user pressing a button for the robot to announce \textit{``Excuse me''} (Fig.~\ref{fig:teaser}--A), rather than directly asking people to move themselves (Fig.~\ref{fig:teaser}--B). Similarly, when an obstacle is blocking the path, delegation means asking the robot to request help from nearby staff (Fig.~\ref{fig:teaser}--C), as opposed to the user attempting to move the obstacle independently.  

Because deciding what action to take requires an understanding of both the environment and the surrounding social context, prior work has identified several categories of queries made by BLV people~\cite{kamikubo2025beyond}. However, many studies in assistive robotics for BLV users, \red{such as \cite{kamikubo2025beyond}}, have relied on Wizard-of-Oz (WOZ) style descriptions rather than machine-generated ones. This approach is limited, as it is unrealistic to assume that a human operator will always be available in real-world deployments. Moreover, participants may recognize when a human is speaking through the robot, which could influence their responses and reduce the authenticity of the interaction.  

Additionally, to evaluate how delegation is reflected in practice, longitudinal studies are necessary. Unlike many assistive robots that are designed to operate autonomously, shared control offers a high level of interactivity, and users may find it difficult to delegate actions to a robot. However, as the novelty effect fades, they may adjust their trust in the robot, refine their willingness to delegate tasks, and develop strategies for interpreting its behavior. Nonetheless, most prior research has examined only short-term or first-time interactions, leaving little understanding of how user perceptions, preferences, and interaction strategies evolve over time. Capturing these longitudinal changes is crucial for designing systems that not only perform well in controlled trials but also support long-term adoption in everyday contexts.

To examine these dynamics, we developed a robot with a shared control interface and conducted a three-week longitudinal study with six blind participants in a real-world museum setting. The system incorporated interactive features to support joint decision-making, including GPT-based environmental descriptions for querying surroundings, obstacle-specific messages explaining why the robot had stopped, and robot-mediated social communication for negotiating crowds or requesting assistance.  

Participants completed navigation tasks such as moving between exhibits, negotiating crowds, approaching lines, and handling physical obstacles. Across the weeks, they engaged with the robot both verbally and physically, making use of its interactive features while also exercising their own judgment about when to act independently. Our study focused in particular on \textit{obstacles at choke points}, \textit{crowds}, and \textit{lines}, as these represent scenarios where autonomous navigation alone is insufficient and some form of social reasoning or interaction is required. Crowds demand negotiation of shared space, obstacles often require assistance from others to be moved, and lines involve subtle reasoning about when and how to engage.  

Our findings show that participants shifted from exploratory to purposeful use of the robot’s functions, increasingly delegating social interactions to the robot in noisy or crowded environments while maintaining independence in quieter settings. This pattern contrasts with prior work done in this area, which has reported that participants preferred to take over interactions when faced with shared control, delegation, or tasks that the robot could not resolve on its own. Our results suggest that under certain real-world conditions, users may instead view robot-led interaction as the more effective option, highlighting the importance of context in shaping delegation preferences.  

Participants also refined their use of environmental descriptions, moving from curiosity-driven activation to targeted requests for context-specific information. 
These behaviors demonstrate that delegation and collaboration are dynamic, shaped by both situational demands and evolving user trust. 
We also observed how participants used the robot to engage with dynamic crowds and real museum visitors, with responses ranging from people stepping aside immediately to cases where requests were misunderstood.  

The main contributions of this paper are:
\vspace{-5pt}
\begin{itemize}
    \item A three-week longitudinal study with six blind participants using a navigation-assistive robot in a real-world museum setting, providing empirical insights into how perceptions of collaboration and social autonomy evolve with repeated interaction.  
    \item Detailed analysis of participants’ evolving strategies for delegation, including when they chose to rely on the robot to initiate social interactions (asking people to move or requesting help) and when they acted independently, as well as how they used environmental descriptions to interpret surroundings and guide decisions.  
    \item Concrete design implications for navigation-assistive robots, including the need for adaptive levels of support, interactive description systems that enable follow-up questions, and flexible communication styles that accommodate both individual preferences and dynamic social environments.  
\end{itemize}

\section{Related Works}

\subsection{Assistive Navigation Systems and Robots for Blind People}

Various navigation systems have been developed to help blind people reach their destinations, typically relying on maps, infrastructure, and localization methods~\cite{bineeth2020blindsurvey,sulaiman2021analysis}.
They have been realized through smartphones~\cite{sato2019navcog3,chen2015blindnavi,kim2016navigating,Kaniwa2024ChitChatGuide}, wearable devices~\cite{lee2014wearable,li2016isana}, cane-extensions~\cite{nasser2020thermalcane}, and more recently, robotic platforms~\cite{lu2021assistive,zhang2023follower,guerreiro2019cabot,kuribayashi2025wanderguide,kayukawa2022HowUsers}.
These robots have gained attention due to their automatic mobility and wayfinding capabilities~\cite{wei2025human}.

Multiple robotic form factors have been explored, including companion-like~\cite{feng2015designing}, cane-like~\cite{ranganeni2023exploring}, quadruped~\cite{zhang2023follower,wang2021navdog,kim2023transforming,doore2024co}, and wheeled designs~\cite{lu2021assistive,zhang2023follower,guerreiro2019cabot}, each offering distinct benefits.
Quadrupeds, for example, can navigate stairs and uneven outdoor terrain~\cite{cai2024navigating}, while many blind users prefer wheeled robots for their stability and lower noise levels~\cite{wang2022can}. One widely deployed example is the suitcase-style \textit{AI Suitcase}~\cite{Takagi2025FieldTrials}.

Even with these advances, navigation robots still face challenges in dense, dynamic environments such as busy public spaces~\cite{Takagi2025FieldTrials}.
This difficulty is often described as the ``\textit{freezing robot problem}’’\cite{wang2023social,trautman2010unfreezing}.
Researchers have explored social navigation planners\cite{wang2023social}, as well as alerting methods to signal pedestrians~\cite{kayukawa2019bbeep}, to mitigate this issue.
More recently, shared control has emerged as a promising way to handle situations a robot cannot resolve autonomously, allowing users to directly shape navigation decisions~\cite{wang2017trends,kuribayashi2023pathfinder,kamikubo2025beyond}.

\red{Building on this line of work, we integrated an enhanced shared-control system that supports both navigation and on-demand situational awareness in challenging social environments. Prior suitcase-style robots provided notifications when being blocked by people and approaching exhibitions ~\cite{cabot}, but they did not allow users to request environmental information when needed. To address this gap, we incorporated a description feature that generates broad environmental summaries rather than only safety-related notifications. We designed a general-purpose prompt that produces descriptions of objects, people, spatial layout, and overall scene context. This approach was inspired by open-ended exploration tools such as Wanderguide~\cite{kuribayashi2025wanderguide}, which offer contextual descriptions when user intent is not yet known.}

\red{We then used this enhanced system to investigate how blind participants practically engage with shared control in situations where the robot alone cannot resolve the environment, focusing on how users choose between acting independently or relying on robot-mediated social communication.}

\subsection{Delegation in Social Interactions}

Although fully automatic robotic navigation may seem ideal for blind individuals, failures can leave users in situations where they cannot obtain assistance~\cite{kamikubo2025beyond}. To address these risks, the shared control paradigm has emerged as a promising interaction method for navigation systems for the blind, as it is known to improve confidence, comfort, and trust~\cite{zhang2023follower,ranganeni2023exploring}. In shared control, user input and robot autonomy are dynamically combined to collaboratively complete navigation tasks.

Within this paradigm, delegation can be understood as a specific form of shared control, where authority is explicitly transferred between the human and the robot. Rather than blending inputs continuously, the user may delegate subtasks, such as obstacle avoidance or route following, to the robot, while retaining high-level decision-making power. This framing highlights the importance of preserving human agency: the robot should enhance, rather than override, the user’s autonomy~\cite{kamikubo2025beyond}. \red{Compared to ~\cite{kamikubo2025beyond} our work contributes new insights by examining delegation in a real-world setting that involves longitudinal exposure, longer routes, more scenarios, and a fully autonomous system rather than a Wizard of Oz approach. This combination provides a richer understanding of how delegation preferences develop over time and how users negotiate control during everyday navigation tasks.}

Prior efforts in the accessibility domain illustrate how such approaches can be applied. For example, shared-control systems have been used to support wheelchair navigation~\cite{seki2000autonomous,subramanian2019eye,perrin2010brain}, and guide robots have enabled blind users to specify proceeding directions at intersections, while the robot provides automated guidance to the next landmark~\cite{kuribayashi2023pathfinder,jain2023want,lacey2000context,hwang2022system}. Other work has focused on exploration support in unfamiliar environments, enabling blind users to flexibly switch between direct control and automated guidance~\cite{kuribayashi2023pathfinder,kuribayashi2025wanderguide}.

These examples reinforce the idea that delegation within shared control is not merely about dividing labor; it is about creating mutual support systems where both the human and the robot actively contribute to successful navigation. Such systems must be designed to preserve decision-making authority while offering reliable assistance, thereby promoting independence without diminishing user control.

\subsection{Longitudinal Studies}

Longitudinal studies allow participants to gain extended exposure to new technologies, providing insights into how systems are adopted, adapted, and sustained over time. Such studies are particularly valuable for moving beyond short-term novelty effects and capturing how interaction patterns evolve in real-world contexts.

Research on large display systems, VR learning environments, and other interactive technologies has shown that novelty effects are strongest immediately after introduction or when modifications are made, often producing an initial surge in engagement followed by stabilization~\cite{koch2018novelty,vr_longitudinal,huang2006longitudinal,gallacher2015novelty}. These findings highlight that short-term trials may misrepresent long-term adoption, since early behaviors often differ substantially from stable usage patterns. Multi-week deployments are therefore essential for understanding how strategies evolve once novelty diminishes and real-world routines emerge. These prior works show the importance of longitudinal approaches in revealing not only how novelty shapes early adoption but also how user behaviors and perceptions stabilize over time.

% \red{Although there is limited work on longitudinal navigation studies with blind participants, prior longitudioanl works in related domains,such as ~\cite{vr_longitudinal}, where partiticpants repeated simiar tasks acorss essions,informed our deisng choices. However, ~\cite{OlderAdultsLongitudional} found that appropriate duration for such studies remains context depending on methodology and is still considered an open question within HRI. 
% }

% Despite this, very few studies have applied longitudinal methods to assistive robots~\cite{wei2025human}, particularly in navigation research. Existing efforts have mainly focused on non-navigation contexts, such as classrooms~\cite{neto2021community}, gaming~\cite{metatla2020robots}, or even non-robotic assistive technologies~\cite{rocha2025exploring}. In navigation assistance, most studies still evaluate systems in short-term trials lasting only a few hours~\cite{wei2025human}. \red{Although some projects develop navigation systems iteratively through hardware updates and deployments in large facilities~\cite{takagi2025field}, individual longitudinal familiarization and adaptation remain underexplored.}

\red{Although there is limited work on longitudinal navigation studies with blind participants, prior longitudinal studies in related domains, such as~\cite{vr_longitudinal}, where participants repeated similar tasks across sessions, informed our design choices. However, as noted in~\cite{OlderAdultsLongitudional}, the appropriate duration for longitudinal studies remains context dependent and is still considered an open methodological question within HRI.}

Despite this, very few studies have applied longitudinal methods to assistive robots~\cite{wei2025human}, particularly in navigation research. Existing efforts have mainly focused on non-navigation contexts, such as classrooms~\cite{neto2021community}, gaming~\cite{metatla2020robots}, or even non-robotic assistive technologies~\cite{rocha2025exploring}. In navigation assistance, most evaluations still rely on short-term trials lasting only a few hours~\cite{wei2025human}. \red{Although some projects develop navigation systems iteratively through hardware updates and deployments in large facilities~\cite{Takagi2025FieldTrials}, individual longitudinal familiarization and adaptation remain underexplored.}

In this work, we address this gap by conducting a \red{three session repeated-exposure study across three weeks} with blind participants using a navigation-assistive robot. Our focus is on how delegation in social interactions evolves over time, going beyond single-session familiarization to investigate how participants balance independence with robot assistance in repeated, real-world use.

\section{Longitudinal Study Methods}
In this work, we conducted a three-week longitudinal study with six blind individuals, each of whom interacted with the navigation robot across three separate weeks. We first describe our robotic system, which adopts a shared control interface, followed by details of the study.

\subsection{Robot Function and Interaction}

\begin{figure*}[t]
    \centering
    \begin{subfigure}[b]{0.45\textwidth}
        \centering
        \includegraphics[width=\textwidth]{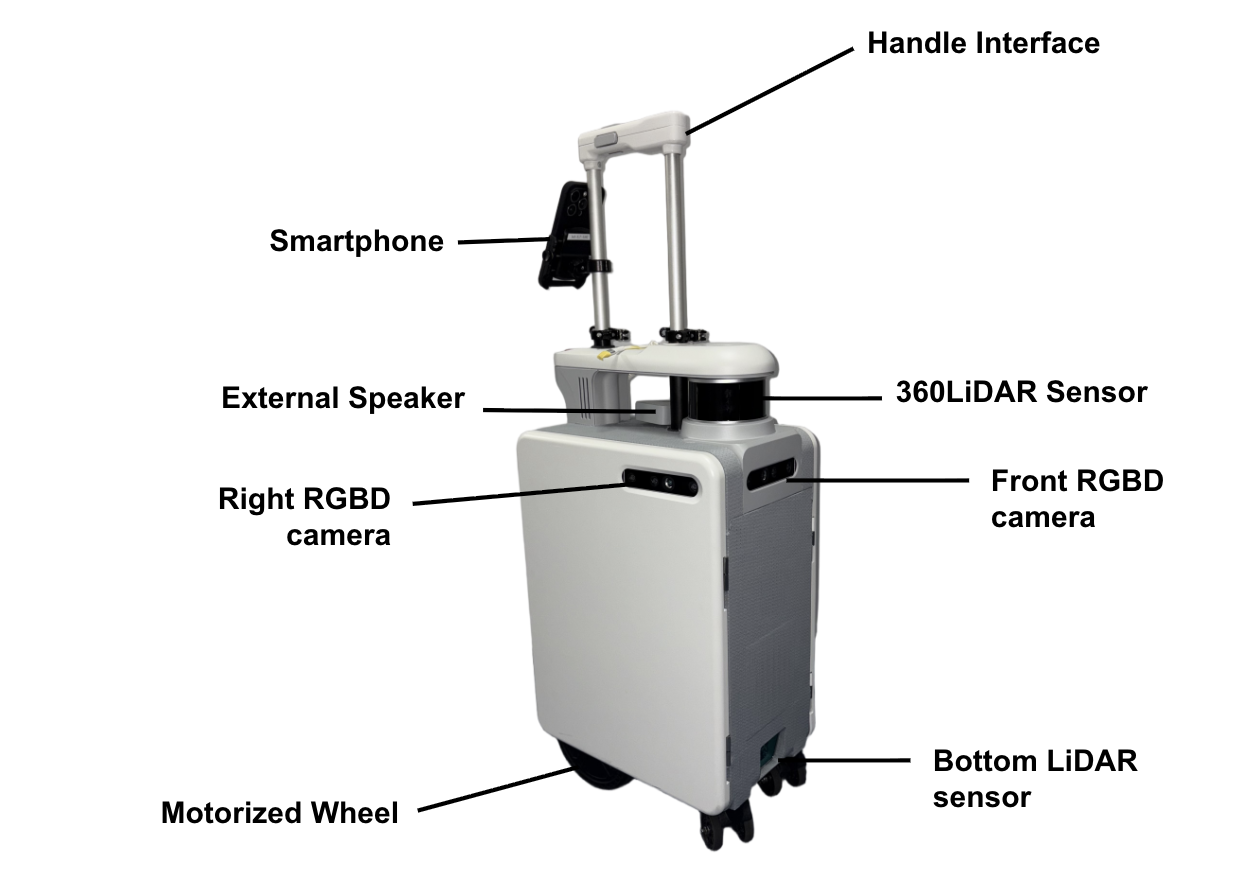}
        \caption{Front view}
        \Description{Annotated photo of a suitcase-shaped mobile robot with sensing and interaction components. Labels point to the handle interface at the top, a mounted smartphone, an external speaker, a 360-degree LiDAR sensor, a front RGBD camera, a right-side RGBD camera, a bottom LiDAR sensor, and motorized wheels at the base.}
        \label{fig:robot_front}
    \end{subfigure}
    \hfill
    \begin{subfigure}[b]{0.45\textwidth}
        \centering
        \includegraphics[width=\textwidth]{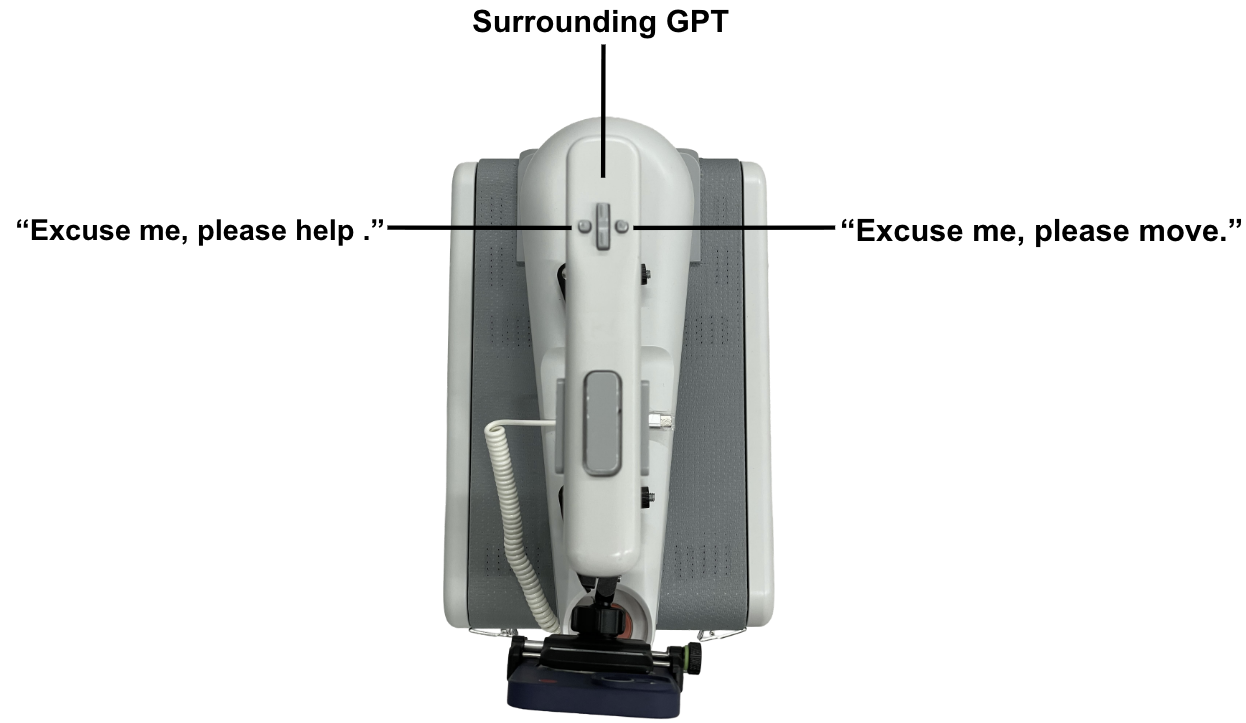}
        \caption{Top-down view}
        \Description{Annotated top-down photo of the handle of the suitcase-shaped mobile robot. There are three buttons towards the top of the robot with labels. Labels point to its sides and top, indicating interaction zones with text prompts: “Excuse me, please help” on the left, “Excuse me, please move” on the right, and “Surrounding GPT” at the top.}
        \label{fig:robot_topdown}
    \end{subfigure}
    \caption{\textbf{Guide Robot.} The suitcase-shaped robot used in the study. It integrates a handle interface, smartphone, and external speaker for user interaction. It includes multimodal sensing through front and right RGBD cameras, a 360 LiDAR, a bottom LiDAR, and motorized wheels for navigation.}
    \label{fig:robot_combined}
\end{figure*}

We used an open-source robotic system~\cite{cabot}, modified to serve as the guide robot for this study (Fig.~\ref{fig:robot_front}). The robot is equipped with LiDAR sensors and three RGB-D cameras (front, left, and right) to support navigation and obstacle detection. Its handle interface integrates a touch sensor and four buttons (top, bottom, left, and right), allowing participants to issue commands and receive haptic feedback. A smartphone mounted on the back of the robot communicated with a neck speaker worn by participants, which delivered GPT-generated obstacle detections and environmental descriptions. An external speaker on the robot was used for preset social phrases when participants delegated social interactions to the robot.  

\subsubsection{Surrounding GPT}

The \textbf{top button} allowed participants to request a description of their surroundings, which we refer to as \textit{``Surrounding GPT''}. Pressing this button captured a frame from the front RGBD camera and submitted it to GPT using the prompt \textit{``Please describe this scene.''}. \red{The prior version of the robot~\cite{cabot} did not provide users with a way to request information about their surroundings. It only delivered notifications about nearby obstacles and approaching exhibits, without accepting any user-initiated queries. To address this limitation, we introduced Surrounding GPT as a method for offering general situational awareness rather than only safety related alerts. We designed a generic prompt that encouraged the model to describe the scene broadly, informed by the principles of Wanderguide ~\cite{kuribayashi2025wanderguide}, which supports exploration by providing open-ended contextual descriptions when user intent is not yet known. This resulted in outputs that included visible objects, people, spatial layout, and higher level semantic impressions of the environment.}

The \textbf{right button} enabled participants to delegate social interaction by having the robot say \textit{``Excuse me, please move.''}, while the \textbf{left button} prompted the robot to vocalize \textit{``Excuse me, please help me.''}. Together, these functions gave participants multiple ways to manage social interactions, supporting both independent action and delegation to the robot depending on the situation. This feature was developed on top of the open sourced robot platform~\cite{cabot}. \red{While the robot’s compact, seamless design allows users to blend naturally into their environment, this advantage also introduces limitations not typically encountered with more visibly identifiable navigation tools such as a white cane or guide dog, since bystanders may not immediately recognize the device or know to make space or offer assistance.}

\subsubsection{Obstacle Detection}

The robot includes a built-in obstacle detection system that classifies obstacles into general categories such as \textit{person}, \textit{object in front}, or \textit{unknown obstacle}. To enrich this capability, we developed an additional layer that provides more detailed contextual information. When one of the three categories is detected, the system captures a frame from the front RGB camera and sends it to GPT-4o along with the following prompt:

\begin{quote}
You are an assistant guiding a person with visual impairment. The robot is currently stopped for one of the following reasons: (1) there is a person in front, (2) it is trying to avoid a person, or (3) it has detected an obstacle. Look at the image and, following the rules below, state a concise one-sentence explanation corresponding to the reason for the stop.

(1) If there is a person in front, or (2) if the robot is trying to avoid a person: state only the approximate number of people, whether they are in a line or a crowd, and whether they are facing toward you or facing away.  
Example: ``There is a crowd ahead with about three people, facing away.''

(3) If the robot has detected an obstacle: briefly state what the obstacle is.  
Example: ``The robot has stopped due to an obstacle. There is a cone ahead.''

Please strictly follow the rules below:
\begin{itemize}
  \item Do not provide any other information or descriptions.
  \item Do not use vague expressions or apologies such as ``I do not know,'' ``I cannot judge,'' or ``I am sorry.''
  \item Do not include opinions or subjectivity; state only objective facts.
  \item Keep the explanation as short and concise as possible.
\end{itemize}
\end{quote}

The prompt was provided in the local language to ensure that the generated output was also in that language.

\subsubsection{Navigation and Interaction}
Participants navigated by holding onto the robot’s handle. Once a destination was set through the Bluetooth-connected smartphone, participants indicated readiness to begin moving by pressing the right button. The robot then autonomously navigated to the destination, following waypoints and avoiding obstacles and other visitors.  

In situations where the robot could not move past people, such as static crowds or physical obstacles, either the robot or the participant needed to initiate a social interaction with those nearby to resolve the problem. \red{Participants were informed that they could either use the robot’s buttons to have the system ask people to move or request assistance from nearby individuals, or they could choose to manage these interactions independently.}

\begin{figure*}[h]
    \centering
    % First image
    \begin{subfigure}[b]{0.3\textwidth}
        \centering
        \includegraphics[width=\textwidth]{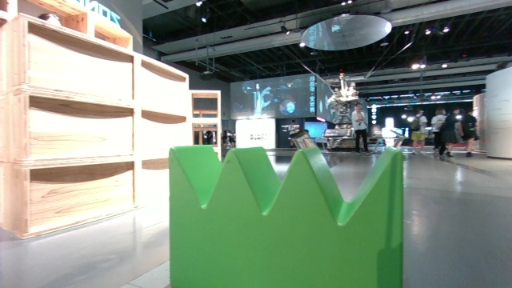}
        \caption{``The robot has stopped due to an obstacle. There is a green obstacle ahead.''}
        \Description{Robot’s front RGBD camera view of a large green foam block from a children’s indoor play exhibit, obstructing part of the walkway.}
        \label{fig:obstacle1}
    \end{subfigure}
    \hfill
    % Second image
    \begin{subfigure}[b]{0.3\textwidth}
        \centering
        \includegraphics[width=\textwidth]{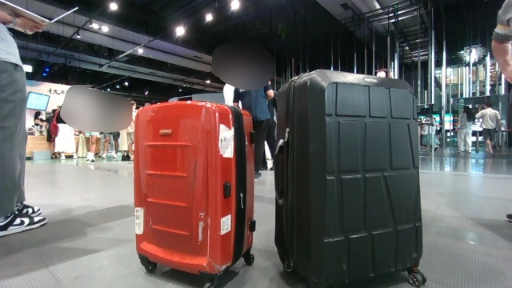}
        \caption{``The robot has stopped due to an obstacle. There are suitcases ahead.''}
        \Description{Robot’s front RGBD camera view of two large suitcases, one red and one black, blocking the walkway.}
        \label{fig:obstacle2}
    \end{subfigure}
    \hfill
    % Third image
    \begin{subfigure}[b]{0.3\textwidth}
        \centering
        \includegraphics[width=\textwidth]{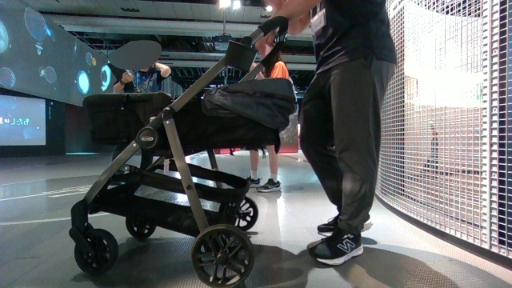}
        \caption{``The robot has stopped due to an obstacle. There is a baby stroller ahead.''}
        \Description{ Robot’s front RGBD camera view of a stroller with a person holding the stroller obstructing the path.}
        \label{fig:obstacle3}
    \end{subfigure}

    \caption{\textbf{Route Object Obstacles.} Examples of object obstacles encountered during navigation with GPT-generated descriptions, as captured by the robot’s front RGBD camera. Week 1 included a green foam block from a children’s play exhibit, Week 2 featured two suitcases, and Week 3 presented a baby cart.}
    \label{fig:obstacles}
\end{figure*}

\subsection{User Study Detail}
Using our robot system, we conducted a three-week study with six blind participants, which was approved by our institutional review board (IRB).  

\subsubsection{Recruitment and Participants}

The study was advertised to the local blind and low-vision community through a mailing list affiliated with our institution. Six participants (three women and three men) between the ages of 18 and 69 were recruited ($M = 42.0$, $SD = 23.6$). Table~\ref{tab:demographics} presents a detailed breakdown of participant demographics, including age, gender, vision level, duration of impairment, navigation tools used, and frequency of independent outings. \red{
Given the difficulty of recruiting blind participants for long-term studies, our system was evaluated with six users.
This sample size aligns with usability research indicating that a single study with approximately five users can reveal about 75\% of usability issues~\cite{number_users,kuribayashi2025wanderguide,kuribayashi2023pathfinder}.
}

\renewcommand{\arraystretch}{1.5}

\begin{table*}[h]
\centering
\caption{\textbf{Participant demographics.}}
\Description{Table summarizing participant demographics. Six participants (P1–P6) are listed with details on age, gender, vision level, impairment duration, navigation tools, and independent outings. Ages range from 18 to 69. Four participants are blind and two have low vision. Impairment duration is typically greater than 10 years, except for P4 (<5 years). Navigation tools include white canes, guide dogs, GPS apps, and BeMyAI. Frequency of independent outings ranges from every day to about once per week.}
\label{tab:demographics}
\renewcommand{\arraystretch}{1.3}
\resizebox{\textwidth}{!}{
\begin{tabular}{lclllll}
\toprule
\textbf{ID} & \textbf{Age} & \textbf{Gender} & \textbf{Vision Level} & \textbf{Impairment Duration} & \textbf{Navigation Tools} & \textbf{Independent Outings} \\
\midrule
P1 & 21 & Man   & Blind & $>$10 years & White cane                   & Every day \\
P2 & 66 & Woman & Blind & $>$10 years & Guide dog, apps              & Every day \\
P3 & 69 & Man   & Blind & $>$10 years & White cane, apps, guide      & Every day (guide 3--4 times/month) \\
P4 & 18 & Woman & Blind & $<$5 years  & White cane, GPS apps         & $\sim$5 times/week \\
P5 & 54 & Man   & Blind & $>$10 years & White cane, BeMyAI           & Every day \\
P6 & 24 & Woman & Low vision & $>$10 years & White cane, apps        & About once/week \\
\bottomrule
\end{tabular}
}
\end{table*}

\subsubsection{Scenarios \& Routes}

\red{We designed each route intentionally to expose participants to a diverse set of navigation challenges. The routes included narrow hallways, long corridors, chokepoints, queues, and crowded areas, allowing us to emulate the types of situations blind travelers commonly encounter in public indoor environments.} 

\red{To guide this design process, we developed a set of principles and guidelines informed by a prior work on social navigation~\cite{SocialNav}. These guidelines emphasized creating situations where users would need to negotiate space, manage congestion, and determine when to rely on the robot for assistance versus when to take direct control.}

\red{We expanded and adapted the routes explored in ~\cite{kamikubo2025beyond}. Based on these guidelines, our routes were significantly longer and more complex than those used in prior work. The repeated-exposure structure of our study allowed participants to experience multiple instances of social navigation patterns across different spaces. This enabled us to build on established navigation patterns while incorporating more complex and extended paths appropriate for longitudinal evaluation. By integrating both structured challenges and open-ended environments, the routes offered a comprehensive testbed for observing delegation behaviors and shared control in realistic conditions.}

% \begin{figure}[t]
%     \centering
%     \includegraphics[width=\linewidth]{figures/routes.pdf}
%     \caption{\textbf{Navigation Routes.} Routes used in the study as mapped on museum floors. Week 1 and Week 3 were conducted on Floor A, while Week 2 was on Floor B. Across the routes, participants encountered a variety of obstacles, including crowds, lines, luggage, and baby carts.}
%     \label{fig:routes}
% \end{figure}

\begin{figure*}[t]
    \centering
    \includegraphics[width=\textwidth]{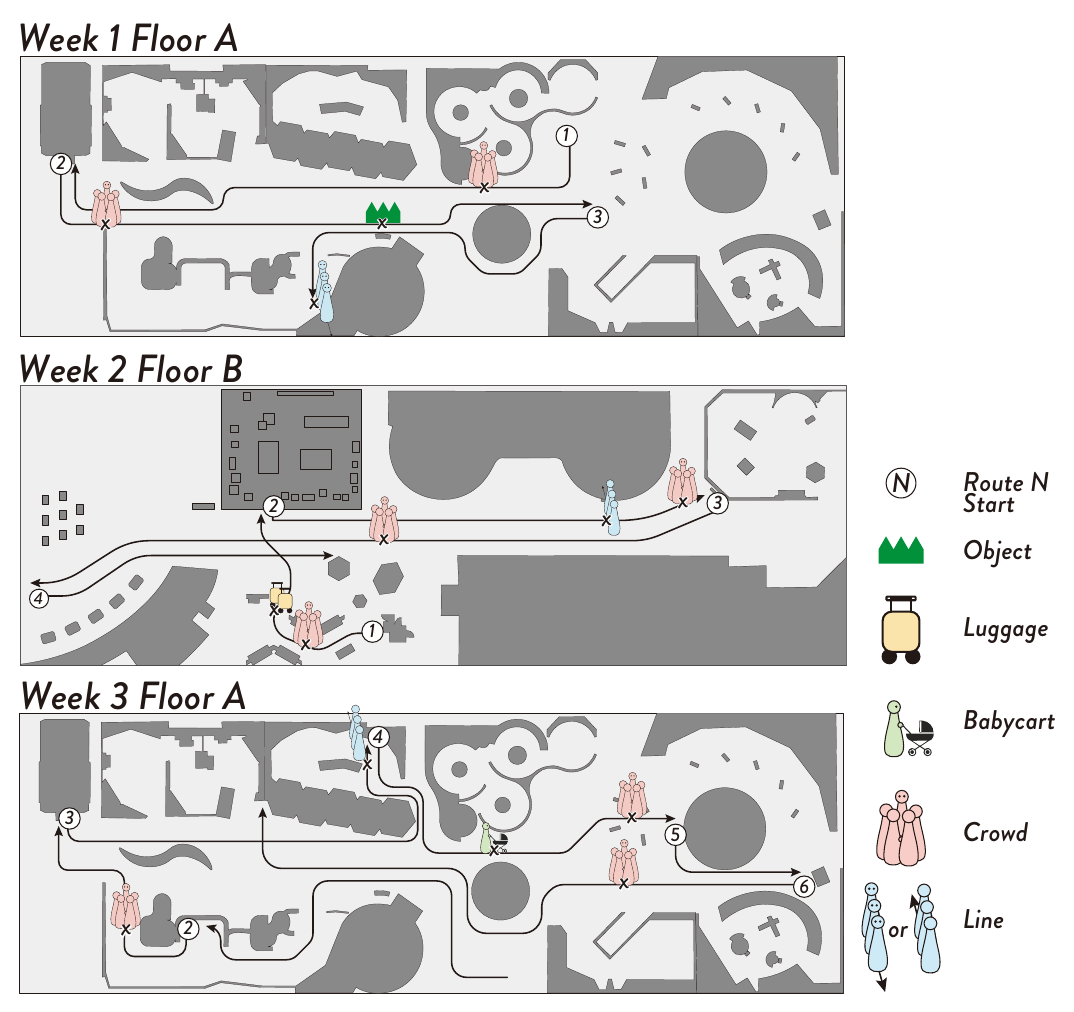}
    \caption{\textbf{Navigation Routes.} Routes used in the study as mapped on museum floors. Week 1 and Week 3 were conducted on Floor A, while Week 2 was on Floor B. Across the routes, participants encountered a variety of obstacles, including crowds, lines, luggage, and baby carts.}
    \Description{Diagram showing the navigation routes used in the study across three weeks. In Week 1 on Floor A, there were three routes: crowds were encountered on routes 1 and 2, an object obstacle on route 2, and a line at the end of route 3. In Week 2 on Floor B, there were four routes: route 1 included a crowd followed by a luggage obstacle, route 2 included a line followed by a crowd, route 3 included a crowd, and route 4 had no obstacles. In Week 3 on Floor A, there were six routes: route 1 had no obstacles, route 2 included a crowd, route 3 included a line, route 4 included a baby cart followed by a crowd, route 5 had no obstacles, and route 6 included a crowd.}
    \label{fig:routes}
\end{figure*}
We designed three distinct scenarios in which participants were required to take an action to address the situation, following the framework established in a prior work~\cite{kamikubo2025beyond}.

\begin{enumerate}
    \item \textbf{Crowd scenario:} A static crowd blocked the path. Participants could either speak directly or press the right button for the robot to say \textit{``Excuse me, please move.''}  
    \item \textbf{Obstacle scenario:} An object blocked the path. Participants could press the left button for the robot to request help or do so themselves.  
    \item \textbf{Line scenario:} A line of people formed at the destination. Participants decided whether to initiate the interaction themselves or delegate it to the robot.  
\end{enumerate}

Fig.~\ref{fig:routes} illustrates the routes and locations where the scenarios were conducted over the three weeks.  
The \textit{crowd scenario} was created by two experimenters who stood close together to form a choke point. They remained stationary unless the participant acted, either by commanding the robot to speak or verbally asking them to step aside.

For the \textit{obstacle scenarios}, three different obstacles were introduced, one per week (Fig.~\ref{fig:obstacles}). In the first week, a simple grass-like foam object was used. \red{This object was taken directly from a museum exhibit to simulate a pop-up exhibition.} In the second week, two suitcases were placed.

In the final week, a baby stroller accompanied by a man looking at his phone was positioned.\red{These suitcase and stroller obstacles were selected because they are common items brought into the museum by visitors and therefore reflect realistic, everyday conditions.} Unless participants attempted to move the obstacle, asked for assistance, or requested the stroller to be moved, the objects remained in place. An experimenter acted as a nearby bystander, ready to assist only when asked. \red{Although the robot was deployed during regular museum hours and participants encountered natural crowds, some obstacles needed to be introduced because museum staff routinely cleared unattended objects. Adding controlled obstacles ensured that all participants experienced comparable navigation challenges in consistent locations, allowing fair comparison while still situating the study within a naturalistic environment.}

The \textit{line scenarios} varied across the three weeks. In the first and third weeks, a line blocked the area near the destination, preventing the robot from reaching it. Because robots often misclassified lines as crowds, participants were required to seek help from surrounding people to interpret the situation and wait patiently. The line moved forward one by one only after the participant recognized it as a line by asking the nearby bystander. In the second week, the line directly blocked the participant’s path, requiring them to ask people in line to move aside so they could pass. Each line consisted of two to three researchers. Importantly, in all weeks, participants also encountered real dynamic crowds of museum visitors, as the study took place during regular museum hours.

\subsubsection{Procedure}
Participants attended three sessions (approximately two hours each), scheduled once per week, and spent about 60 minutes per session walking routes with the robot. The overall procedure remained consistent across the three weeks, with only minor details varying. \red{Although there is limited work on longitudinal navigation studies with blind participants, our approach was informed by longitudinal studies in related domains where participants repeated similar tasks across sessions~\cite{vr_longitudinal,repeated}, and the appropriate duration for such studies remains context dependent and an open question within HRI~\cite{OlderAdultsLongitudional}. Prior works show that stable conditions help reduce novelty effects and allow participants to focus on using the system rather than adapting to new elements each week, leading to more meaningful feedback about the system itself.}

In the first session, participants completed consent forms and provided demographic information. Each week began with a training session. During Week ~1, training lasted about 30 minutes to ensure participants became familiar with all robot functions. This included practicing walking with the handle, using the buttons, and experiencing each of the three scenarios while trying all possible actions. In Weeks ~2 and ~3, participants completed a shorter refresher training session to help them recall the robot’s functions.  

After training, participants walked three predetermined routes during regular museum hours, encountering both staged and natural obstacles, as shown in Fig.~\ref{fig:routes}. \red{The labels refer to the planned obstacles and crowd scenarios introduced by the researchers.} Each session concluded with a brief post-session interview about their experiences and reflections, which will be described in the following section.

\red{Participants were not informed that certain scenarios would be performed by actors, and were asked to respond as they naturally would in a real-world setting. Actors were trained to follow consistent scripts for obstacle placement and crowd behavior, ensuring comparability across participants. The robot’s spoken output from the neck speaker was generally only audible to individuals standing very close to the participant, meaning that surrounding visitors were unlikely to hear or react to the robot’s requests unless they were directly in the interaction space. This helped preserve the naturalistic feel of the environment while still allowing us to study how participants navigated semi-controlled situations.}

\subsubsection{Post-Session Interviews}

At the end of each session, participants completed a short interview combining standardized survey measures and study-specific questions. The survey included the Robotic Social Attributes Scale (RoSAS)~\cite{RoSAS} and five Likert-scale items developed for this study, each followed by an open-ended prompt to capture participants’ reasoning.  

The RoSAS was used to assess participants’ perceptions of the robot’s social attributes, with emphasis on Factor~2 (Competence) and Factor~3 (Discomfort). Participants rated each attribute on a 7-point Likert scale (1 = strongly disagree, 7 = strongly agree):  
\begin{itemize}
    \item \textbf{Competence-related:} capable, responsive, interactive, reliable, competent, knowledgeable  
    \item \textbf{Discomfort-related:} scary, strange, awkward, aggressive, dangerous  
\end{itemize}

Study-specific Likert items addressed delegation and environmental descriptions:  
\begin{enumerate}
    \item When the robot was stopped, I thought it was better for the robot (rather than myself) to ask people to move.  
    \item When the robot was stopped, I thought it was better for the robot (rather than myself) to ask for help.  
    \item I was able to understand my surroundings using the robot’s descriptions.  
    \item I thought the way the robot explained its surroundings was appropriate.  
    \item Even in difficult situations, I was able to confidently decide how to act.  
\end{enumerate}

In weeks ~2 and~3, participants were reminded of their earlier responses and asked if their views had changed. In the final week, they reflected on their overall experience and discussed how their perceptions of collaboration with the robot evolved over the three weeks.

\section{Results}

\subsection{Delegating Social Interaction to the Robot}
Overall, participants increasingly delegated social interactions to the robot over time, though individual trajectories varied, with some showing steady growth and others fluctuating across weeks. 

\begin{table}[h]
\centering
\caption{\textbf{Delegation of Social Communication.} Percentage of interactions in which participants delegated communication tasks (such as asking for help or requesting others to move) to the robot across three weeks. P2–P5 showed consistently high delegation, while P1 and P6 displayed more variability, with P6 gradually increasing reliance on the robot over time. These patterns highlight both stable and evolving strategies of delegation across participants.}
\Description{Table showing the percentage of interactions in which participants delegated social communication to the robot across three weeks. Delegation rates varied by participant and week. P1 increased delegation in Week 2 and decreased in Week 3. P2 consistently showed high delegation rates. P3 and P4 showed high or full delegation across all weeks. P5 increased delegation over time, reaching full delegation by Week 2. P6 showed low delegation initially and increased in Week 3.}
\label{tab:robot_social_use}
\renewcommand{\arraystretch}{1.3}
\begin{tabular}{lccc}
\toprule
\textbf{Participant} & \textbf{Week 1} & \textbf{Week 2} & \textbf{Week 3} \\
\midrule
P1 & 0.0\%   & 57.1\% & 12.5\% \\
P2 & 83.3\%  & 87.5\% & 92.6\% \\
P3 & 85.7\%  & 100\%  & 100\%  \\
P4 & 100\%   & 100\%  & 100\%  \\
P5 & 20.0\%  & 100\%  & 100\%  \\
P6 & 0.0\%   & 20.0\% & 54.5\% \\
\bottomrule
\end{tabular}
\end{table}

Table~\ref{tab:robot_social_use} shows the percentage of interactions in which participants chose to delegate social communication to the robot rather than speaking themselves. \red{This percentage was calculated by dividing the number of times a participant delegated the social interaction by the total number of times they were alerted to an obstacle.}

Participants who began with low levels of delegation (P1, P5, P6) generally increased their reliance on the robot over time, while those who started high (P2, P3, P4) maintained consistently high levels across all three weeks.  \red{Participants showed similar delegation behaviors in both actor-created scenarios and naturally occurring situations (shown in Table ~\ref{tab:natural_obstacles}).}

For P1, delegation rose from none in Week~1 to 57.1\% in Week~2, before falling back to 12.5\% in Week~3. This fluctuation reflects his preference to act independently when possible, while still relying on the robot in particularly noisy or crowded contexts. By contrast, P2 maintained high delegation rates throughout (83.3\%--92.6\%), consistent with her reasoning that the robot’s voice carried better in noisy environments:  
\begin{quote}
``At first, I thought I should say it myself and I did, but it felt like people weren’t really hearing me. With the robot, I felt like people were actually paying attention.'' (P2, Week~3)
\end{quote}

Similarly, P3 consistently delegated (85.7\%--100\%), noting that the robot’s synthetic voice, while not natural, was often more effective at drawing attention than his own:  
\begin{quote}
``I felt that the robot’s voice carried better than my own. Especially in situations with lots of people and a noisy atmosphere, I thought the robot’s voice was more likely to be heard than if I spoke myself.'' (P3, Week~2)
\end{quote}

P4 delegated 100\% of the time across all weeks, which aligned with her self-described reluctance to directly ask for help in public situations:  
\begin{quote}
``It’s just my personality, but I’m not very good at calling out to people myself. For example, when I’m outside and get lost, I’m not the type who can easily say, `Excuse me, could you help me?' In those situations, if the robot had that function, it would really help.'' (P4, Week~1)
\end{quote}

P5 showed one of the most dramatic increases, rising from 20\% in Week~1 to 100\% in Weeks~2 and 3. This trajectory suggests growing trust and comfort in letting the robot handle social interactions on his behalf.  

Finally, P6 demonstrated a gradual increase, moving from 0\% in Week~1 to 54.5\% in Week~3. She explained that when she was uncertain whether people were nearby, the robot’s voice provided a more reliable and less risky way to initiate interaction.  

These trajectories show how quantitative patterns in delegation were closely tied to participants’ lived experiences. Some participants, like P1 and P6, delegated selectively based on situational demands such as noise or uncertainty, while others, like P2, P3, and P4, consistently relied on the robot to overcome personal or environmental barriers to social interaction.

\subsection{Use of Environmental Descriptions}

Table~\ref{tab:surrounding_gpt} shows how often participants activated the ``Surrounding GPT'' function across the three weeks of the study. \red{The numbers shown reflect the number of times each participant requested information about their surroundings by pressing the top button.} Use varied across participants due to individual differences, but overall patterns followed a trajectory from exploration to purposeful engagement. Participants who began with little or no use (P1) gradually incorporated the function into their routines, while others (P6) relied on it heavily and consistently. Some, such as P2 and P4, experimented extensively in Week~2 before becoming more selective in Week~3, suggesting a shift from curiosity-driven exploration to targeted, experience-informed use. \red{P2 and P4 also had increased exposure to obstacle-related GPT triggers during later sessions, which may have influenced their patterns of use.}

\begin{table}[h]
\centering
\caption{\textbf{Use of Surrounding GPT.} Number of times participants triggered the ``Surrounding GPT'' function across three weeks. Usage patterns varied widely: P2 and P4 peaked in Week~2, P6 consistently used the function most, while P5 discontinued after Week~1. These patterns reflect a broader shift from exploratory use to more purposeful engagement over time.}
\Description{Table showing the number of times participants triggered the “Surrounding GPT” function across three weeks. P1 increased from 0 to 2 uses by Weeks 2 and 3. P2 rose from 1 to 8 in Week 2, then dropped to 4. P3 used it 2 times in Week 1, 0 in Week 2, and 3 in Week 3. P4 peaked at 8 in Week 2 but only 1 in Week 3. P5 used it twice in Week 1 and not again. P6 used it most frequently, between 5 and 7 times each week.}
\label{tab:surrounding_gpt}
\renewcommand{\arraystretch}{1.3}
\begin{tabular}{lccc}
\toprule

\textbf{Participant} & \textbf{Week 1} & \textbf{Week 2} & \textbf{Week 3} \\
\midrule
P1 & 0 & 2 & 2 \\
P2 & 1 & 8 & 4 \\
P3 & 2 & 0 & 3 \\
P4 & 2 & 8 & 1 \\
P5 & 2 & 0 & 0 \\
P6 & 7 & 5 & 6 \\
\bottomrule
\end{tabular}
\end{table}

For instance, P6 was the most consistent user, activating the function 7, 5, and 6 times across the three weeks. Her comments reflected this steady reliance:  
\begin{quote}
``While walking, I often wanted a simple overview of what kinds of exhibits were around me, so I pressed the top button to hear the situation explained.'' (P6, Week~3)
\end{quote}

P1, by contrast, never used the feature in Week~1 but incorporated it twice per session in Weeks~2 and 3, showing how initially hesitant users grew to see value in the descriptions. Meanwhile, P2’s activations rose sharply from one in Week~1 to eight in Week~2 before dropping to four in Week~3. She explained that her interest shifted from general crowd awareness to exhibit information:  
\begin{quote}
``When I’m walking through an area, I want to know the situation with the people around me. But when I want to know what kind of place it is, then it’s the exhibit information that I’m interested in.'' (P2, Week~3)
\end{quote}

Several participants also used the descriptions strategically when the robot repeatedly announced that a crowd was blocking its path. In these situations, they sought additional context via Surrounding GPT before deciding whether to act themselves or delegate to the robot. \red{Participants explained that when they received multiple obstacle alerts in quick succession, often due to a continuous flow of dynamic crowds, they activated Surrounding GPT to understand what was happening around them before choosing how to respond. For example, P6 in Week~3 frequently requested a description just before instructing the robot to ask the crowd to move. }

Overall, participants were satisfied with the GPT-based descriptions, appreciating that they were concise and actionable. \red{At the same time, preferences diverged: some felt the outputs were occasionally too broad and expressed interest in a more interactive system that would allow follow-up questions, while others valued the broader overviews as a way of situating themselves in the environment \cite{kuribayashi2025wanderguide}. }

These results show that GPT-generated descriptions played a complementary role in collaboration: they not only provided situational awareness but also influenced participants’ decision-making about when and how to intervene, thereby supporting calibration of shared responsibility between user and robot.

\subsection{Responses to Obstacle-Related GPT Triggers}

% \begin{table}[h]
% \centering
% \caption{\textbf{Obstacle-Related GPT Usage and Actions.} Number of obstacle-related GPT triggers and corresponding action percentages across three weeks. Usage varied widely: P2 and P6 showed the highest number of triggers by Week 3, while action percentages were generally moderate, with P6 showing the highest early adoption but a decline over time, and P2, P4, and P5 reaching higher action rates by Week 3. These findings suggest that participants interpreted the robot’s signals and intervened only when it appeared stuck.}
% \label{tab:obstacle_action_compressed}
% \renewcommand{\arraystretch}{1.3}
% \begin{tabular}{lcccccc}
% \toprule
% \textbf{Participant} & \multicolumn{2}{c}{\textbf{Week 1}} & \multicolumn{2}{c}{\textbf{Week 2}} & \multicolumn{2}{c}{\textbf{Week 3}} \\
% \cmidrule(lr){2-3} \cmidrule(lr){4-5} \cmidrule(lr){6-7}
%  & \textbf{Obstacle GPT} & \textbf{Action \%} & \textbf{Obstacle GPT} & \textbf{Action \%} & \textbf{Obstacle GPT} & \textbf{Action \%} \\
% \midrule
% P1 & 15 & 27\% & 29 & 24\% & 19 & 42\% \\
% P2 & 10 & 60\% & 18 & 44\% & 47 & 57\% \\
% P3 & 12 & 58\% & 11 & 55\% & 18 & 44\% \\
% P4 & 12 & 33\% & 16 & 44\% & 14 & 57\% \\
% P5 & 11 & 45\% & 15 & 40\% & 14 & 57\% \\
% P6 & 7  & 71\% & 12 & 42\% & 28 & 39\% \\
% \bottomrule
% \end{tabular}
% \end{table}

\begin{table*}[t]
\centering
\caption{\textbf{Obstacle-Related GPT Usage and Actions.} Number of obstacle-related GPT triggers and corresponding action percentages across three weeks. Usage varied widely: P2 and P6 showed the highest number of triggers by Week 3, while action percentages were generally moderate, with P6 showing the highest early adoption but a decline over time, and P2, P4, and P5 reaching higher action rates by Week 3. These findings suggest that participants interpreted the robot’s signals and intervened only when it appeared stuck.}
\Description{\Description{Table showing the number of obstacle-related GPT triggers and action percentages across three weeks for each participant. P1 triggered GPT 15, 29, and 19 times with action rates of 27\%, 24\%, and 42\%. P2 triggered 10, 18, and 47 times with action rates of 60\%, 44\%, and 57\%. P3 triggered 12, 11, and 18 times with action rates of 58\%, 55\%, and 44\%. P4 triggered 12, 16, and 14 times with action rates of 33\%, 44\%, and 57\%. P5 triggered 11, 15, and 14 times with action rates of 45\%, 40\%, and 57\%. P6 triggered 7, 12, and 28 times with action rates of 71\%, 42\%, and 39\%.}}
\label{tab:obstacle_action_compressed}
\renewcommand{\arraystretch}{1.3}
\begin{tabular}{lcccccc}
\toprule
\textbf{Participant} & \multicolumn{2}{c}{\textbf{Week 1}} & \multicolumn{2}{c}{\textbf{Week 2}} & \multicolumn{2}{c}{\textbf{Week 3}} \\
\cmidrule(lr){2-3} \cmidrule(lr){4-5} \cmidrule(lr){6-7}
 & \textbf{Obstacle GPT} & \textbf{Action \%} & \textbf{Obstacle GPT} & \textbf{Action \%} & \textbf{Obstacle GPT} & \textbf{Action \%} \\
\midrule
P1 & 15 & 27\% & 29 & 24\% & 19 & 42\% \\
P2 & 10 & 60\% & 18 & 44\% & 47 & 57\% \\
P3 & 12 & 58\% & 11 & 55\% & 18 & 44\% \\
P4 & 12 & 33\% & 16 & 44\% & 14 & 57\% \\
P5 & 11 & 45\% & 15 & 40\% & 14 & 57\% \\
P6 & 7  & 71\% & 12 & 42\% & 28 & 39\% \\
\bottomrule
\end{tabular}
\end{table*}

% \begin{table}[t]
%     \centering
%     \caption{\red{Natural occurrences of obstacle alerts and instances where participants became stuck or chose to trigger an interaction. For each week, \textit{GPT} counts the number of times Obstacle GPT was triggered by naturally occurring obstacles or crowds, and \textit{Action} counts the number of times participants either remained stuck for a long time or decided they could not wait and triggered a social interaction.}}
%     \label{tab:natural_obstacles}
%     \renewcommand{\arraystretch}{1.2}
%     \begin{tabular}{lcccccc}
%         \toprule
%         \textbf{Participant} 
%         & \multicolumn{2}{c}{\textbf{Week 1}} 
%         & \multicolumn{2}{c}{\textbf{Week 2}} 
%         & \multicolumn{2}{c}{\textbf{Week 3}} \\
%         \cmidrule(lr){2-3} \cmidrule(lr){4-5} \cmidrule(lr){6-7}
%         & \textbf{GPT} & \textbf{Action} 
%         & \textbf{GPT} & \textbf{Action} 
%         & \textbf{GPT} & \textbf{Action} \\
%         \midrule
%         P1 &  9 & 1 & 13 & 1 & 13 & 2 \\
%         P2 &  6 & 0 & 15 & 2 & 42 & 15 \\
%         P3 &  7 & 2 &  3 & 1 & 15 & 1 \\
%         P4 &  6 & 0 &  6 & 1 &  8 & 2 \\
%         P5 &  3 & 1 &  6 & 1 &  7 & 2 \\
%         P6 &  3 & 0 &  7 & 0 & 21 & 7 \\
%         \bottomrule
%     \end{tabular}
% \end{table}

\begin{table*}[t]
    \centering
    \caption{Natural occurrences of obstacle alerts and instances where participants became stuck or chose to trigger an interaction. For each week, \textit{GPT} counts the number of times Obstacle GPT was triggered by naturally occurring obstacles or crowds, and \textit{Action} counts the number of times participants either remained stuck for a long time or decided they could not wait and triggered a social interaction.}
    \Description{ Table showing the number of natural and stuck events for six participants across three weeks. For Week 1, P1 recorded 9 natural and 1 stuck event. P2 had 6 natural and 0 stuck. P3 had 7 natural and 2 stuck. P4 had 6 natural and 0 stuck. P5 had 3 natural and 1 stuck. P6 had 3 natural and 0 stuck. For Week 2, P1 had 13 natural and 1 stuck event. P2 recorded 15 natural and 2 stuck. P3 had 3 natural and 1 stuck. P4 reported 6 natural and 1 stuck. P5 had 6 natural and 1 stuck. P6 had 7 natural and 0 stuck. For Week 3, P1 had 13 natural and 2 stuck events. P2 had 42 natural and 15 stuck. P3 had 15 natural and 1 stuck. P4 recorded 8 natural and 2 stuck. P5 had 7 natural and 2 stuck. P6 had 21 natural and 7 stuck. The table highlights how each participant’s natural and stuck counts change week by week.
}
    \label{tab:natural_obstacles}
    \renewcommand{\arraystretch}{1.2}
    \begin{tabular}{lcccccc}
        \toprule
        \textbf{Participant} 
        & \multicolumn{2}{c}{\textbf{Week 1}} 
        & \multicolumn{2}{c}{\textbf{Week 2}} 
        & \multicolumn{2}{c}{\textbf{Week 3}} \\
        \cmidrule(lr){2-3} \cmidrule(lr){4-5} \cmidrule(lr){6-7}
        & \textbf{GPT} & \textbf{Action} 
        & \textbf{GPT} & \textbf{Action} 
        & \textbf{GPT} & \textbf{Action} \\
        \midrule
        P1 &  9 & 1 & 13 & 1 & 13 & 2 \\
        P2 &  6 & 0 & 15 & 2 & 42 & 15 \\
        P3 &  7 & 2 &  3 & 1 & 15 & 1 \\
        P4 &  6 & 0 &  6 & 1 &  8 & 2 \\
        P5 &  3 & 1 &  6 & 1 &  7 & 2 \\
        P6 &  3 & 0 &  7 & 0 & 21 & 7 \\
        \bottomrule
    \end{tabular}
\end{table*}

Table~\ref{tab:obstacle_action_compressed} shows the number of obstacle-related GPT triggers and the percentage of those instances in which participants took action, either by using the robot or by acting themselves. \red{To complement the percentage-based analysis in Table~\ref{tab:obstacle_action_compressed}, we additionally report the raw counts of naturally occurring obstacles and instances where participants became truly stuck in Table~\ref{tab:natural_obstacles}. This table shows, for each participant and week, how many of the total Obstacle GPT messages were triggered by real environmental conditions such as dynamic crowds, stationary groups near exhibits, or bottleneck areas within the museum. It also shows how many times participants were unable to proceed for an extended period and either waited or initiated a social interaction. During trials with high crowd densities, we observe more frequent obstacle GPT calls such as in P2 Week~3, and P6 Week~3.}

Overall, participants did not act on every alert, but instead developed strategies to decide when intervention was warranted. For example, P1 encountered 29 obstacle triggers in Week~2 but acted in only 24\% of cases, often waiting to see if the robot would resolve the situation itself. By Week~3, although he encountered fewer triggers (19), his action rate rose to 42\%, reflecting greater willingness to step in when the robot stalled. Similarly, P2’s triggers rose sharply from 10 in Week~1 to 47 in Week~3, yet her action percentage remained stable (60\% to 57\%). She explained that she often interpreted the robot’s small corrective motions—such as backing up or shifting slightly—as signs it was still searching for a path, and therefore chose to wait:  
\begin{quote}
``When the robot stopped and moved a bit this way and that, I thought it was searching—looking for an open path—so I watched it that way.'' (P2, Week~3)
\end{quote}

P6 showed the opposite trajectory: she acted on 71\% of seven triggers in Week~1, but her intervention rate dropped to 39\% when she faced 28 triggers in Week~3. This suggests that as exposure increased, she became more selective about when to intervene, relying on the robot unless it appeared fully stuck. P4 described a similar decision-making process, noting that uncertainty about nearby people often prompted her to act after waiting through multiple alerts:  
\begin{quote}
``When it was moving in a way that seemed a bit uncertain—like stopping and then moving again—or when it seemed like there might be people nearby but I wasn’t sure, those were the times I used it.'' (P4, Week~3)
\end{quote}

Because crowd density varied by day and time, the number of obstacle triggers was not consistent across the sessions, and in some cases repeated alerts were generated for a single situation. This inflated the counts without necessarily prompting more actions.

These findings show that participants learned to interpret the robot’s signals and movements in order to calibrate their own involvement. Rather than treating every obstacle alert as a call to action, they assessed whether the robot was still attempting to resolve the situation or truly needed assistance, highlighting the importance of making the robot’s progress and state more transparent.

\subsection{Learning and Adaptation Over Time}

Over the course of the three weeks, participants noted changes in their understanding of the robot and its functions. They described how they began to anticipate and interpret the robot’s movements as well as adapt their use of its interactive features.  

P2 mentioned that when the robot stopped, she learned to wait before expecting it to move again. She explained that when the robot stopped and made small corrective motions, she gradually recognized that it was searching for a way around an obstacle:  
\begin{quote}
``It was the way the robot moved when there were people around. The small movements, like backing up a little, moving forward, backing up again, turning left, and trying different directions— I gradually came to understand that it was searching for a path.'' (P2, Week 2)
\end{quote}

P1 similarly reported that he was able to anticipate the robot’s movements and speed, particularly when it backed up.  
In addition to learning how to interpret the robot’s behavior, participants also changed how they approached delegating tasks to it. P2 reflected that in the first week she pressed the buttons simply to try them out, but by the final week she was making decisions based on accumulated experience.  

P1 initially resisted delegating to the robot, emphasizing his desire to act independently. In the first week, he reported relying only on himself:  
\begin{quote}
``I think part of it comes from my personality, but for me it was like... from the start, my way of thinking has always been that—even though I can’t see—I’ve always wanted to do the same things as sighted people. Ever since I was little, I’ve felt that way. So since I can call out to people myself, I don’t really feel the need to rely on the robot if it’s something I can just do on my own. Because of that, I think I ended up doing all the calling out myself.'' (P1, Week 1)
\end{quote}

By the second week, however, P1 began using the buttons to delegate to the robot in noisy and crowded settings:  
\begin{quote}
``The reason was that it was crowded today, with a lot of noise around, so my voice often didn’t reach people.'' (P1, Week 2)
\end{quote}

P2 and P3 expressed similar progressions, shifting from initial reliance on themselves to greater use of the robot in situations where their own voices were less effective.  

P6 explained that in the first week she was so accustomed to asking people to move on her own that she did not think to use the robot. In later weeks, however, she began to delegate more to the robot:  
\begin{quote}
``When it was noisy, I felt it was easier for people to notice and help because the robot spoke louder than my own voice.'' (P6, Week 2)
\end{quote}

P5 also described a similar shift. He noted that he preferred using his own voice when only a few people (two or three) were present, but found it difficult in larger crowds or outdoor environments. By the second and third weeks, he reported completely delegating to the robot, explaining that he had grown more accustomed to it, trusted it more, and preferred to let the robot manage these interactions.  

As participants became more familiar with the robot, their use of environmental descriptions also changed. For example, P2 reported using the description button at first because she, in her words, ``just wanted to try it out.''

\subsection{Study-Specific Likert Questions and RoSAS}

Participants’ responses to the four study-specific Likert-scale questions revealed clear trends over the course of the three weeks.  

\begin{figure*}[t]
\centering
\begin{minipage}{0.45\textwidth}
  \centering
  \includegraphics[width=\linewidth]{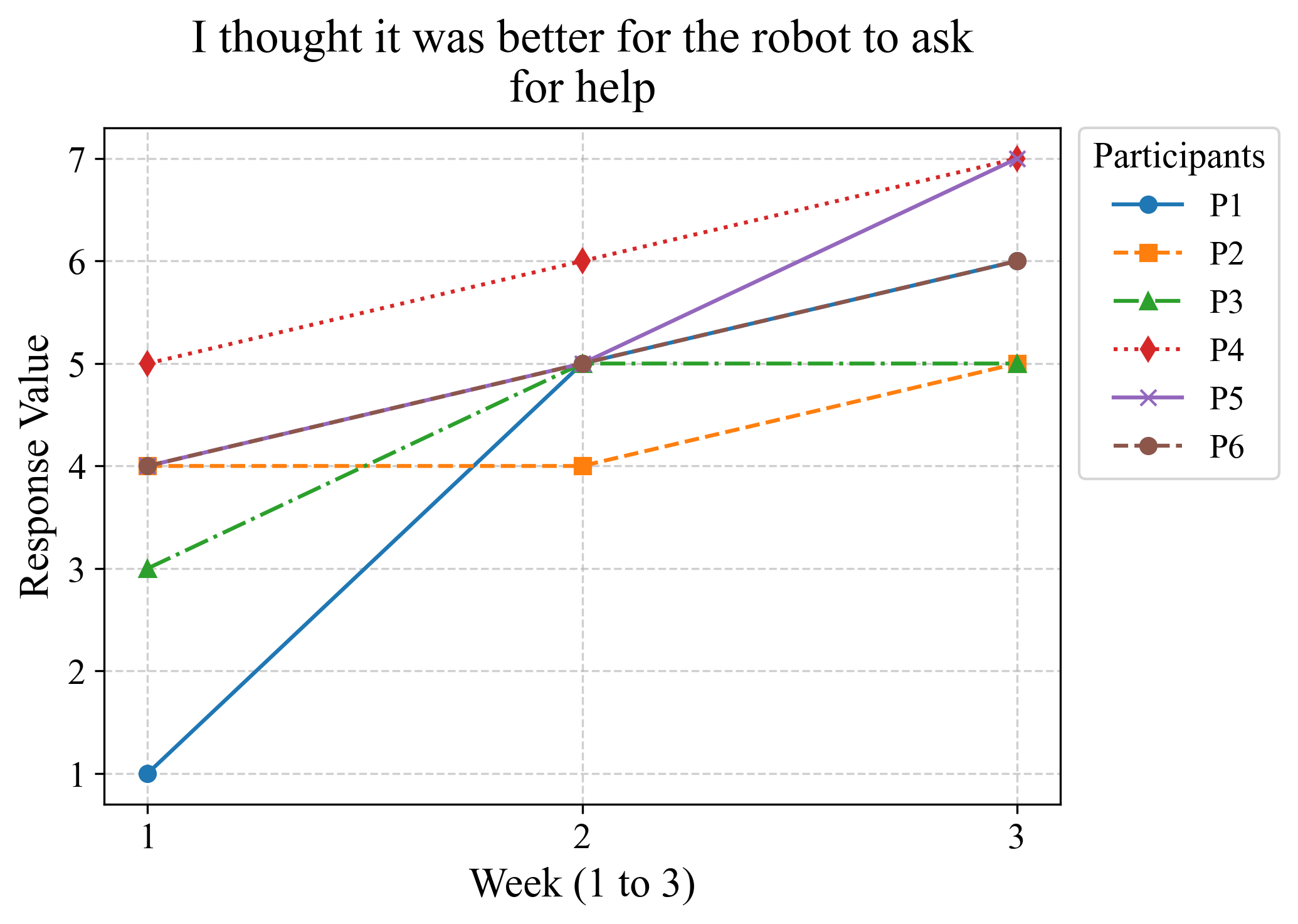}
  \caption{\textbf{Ratings for Statement (1): Preference for the Robot to Ask for Help.} Participant responses over three weeks showed a general upward trend, with P1, P4, and P5 exhibiting the strongest increases, indicating growing acceptance of the robot actively requesting assistance.}
  \label{fig:help_ratings}
  \Description{ Line chart showing participant responses over three weeks to the statement “I thought it was better for the robot to ask for help.” Participant P1 increases from 1 to 6, P2 rises slightly from 4 to 5, P3 increases from 3 to 5, P4 increases steadily from 5 to 7, P5 increases from 4 to 7, and P6 increases from 4 to 6.}
\end{minipage}
\hfill
\begin{minipage}{0.45\textwidth}
  \centering
  \includegraphics[width=\linewidth]{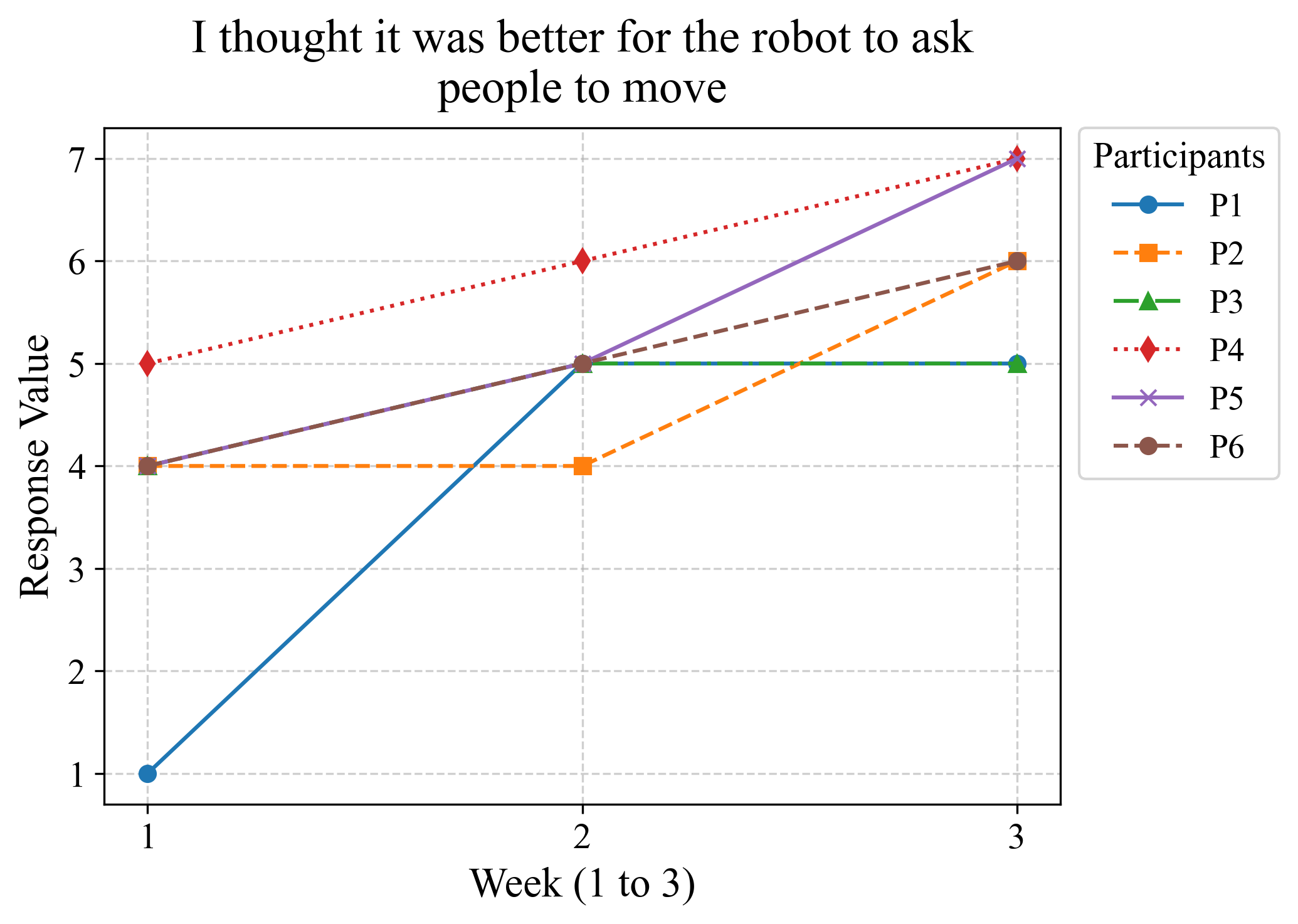}
  \caption{\textbf{Ratings for Statement (2): Preference for the Robot to Ask People to Move.} Participant responses across three weeks showed steadily increasing agreement, with P1, P2, P4, and P5 exhibiting the largest gains, reflecting growing acceptance of robots directly requesting people to clear the way.}
  \Description{Line chart showing participant responses across Weeks 1–3 to the statement “I thought it was better for the robot to ask people to move.” Participant P1 rises sharply from 1 in Week 1 to 5 in Week 2, then to 7 in Week 3. P2 stays at 4 through Week 2, then increases to 6 in Week 3. P3 rises from 3 to 5 by Week 2 and stays flat. P4 increases steadily from 5 to 7. P5 rises from 4 to 7. P6 increases gradually from 4 to 6.}
  \label{fig:move_ratings}
\end{minipage}
\end{figure*}

For statements (1) and (2), which asked whether participants preferred the robot to initiate social interactions (i.e., asking people to move or asking for help) instead of doing so themselves, we observed a shift toward stronger agreement over time (Fig.~\ref{fig:help_ratings} and Fig.~\ref{fig:move_ratings}). \red{Linear mixed–effects analyses for Statements~1 and~2 showed clear positive effects of Week on participants’ ratings. Week was a highly significant predictor for both items (Statement~1: $F(2,10) = 13.57$, $p = .0016$; Statement~2: $F(2,10) = 13.21$, $p = .0014$), indicating strong increases in participants’ comfort with the robot initiating social interactions over time.}

% \begin{figure}[h]
%     \centering
%     % First pair of figures (Help and Move)
%     \begin{minipage}[t]{0.48\textwidth}
%         \centering
%         \includegraphics[width=\linewidth]{images/I_thought_it_was_better_for_the_robot_to_ask_for_help.png}
%         \caption{\textbf{Ratings for Statement (1): Preference for the Robot to Ask for Help.} Participant responses over three weeks showed a general upward trend, with P1, P4, and P5 exhibiting the strongest increases, indicating growing acceptance of the robot actively requesting assistance.}
%         \label{fig:help_ratings}
%     \end{minipage} \hfill
%     \begin{minipage}[t]{0.48\textwidth}
%         \centering
%         \includegraphics[width=\linewidth]{images/I_thought_it_was_better_for_the_robot_to_ask_people_to_move.png}
%         \caption{\textbf{Ratings for Statement (2): Preference for the Robot to Ask People to Move.} Participant responses across three weeks showed steadily increasing agreement, with P1, P2, P4, and P5 exhibiting the largest gains, reflecting growing acceptance of robots directly requesting people to clear the way.}
%         \label{fig:move_ratings}
%     \end{minipage}
% \end{figure}

\begin{figure*}[t]
\centering
\begin{minipage}{0.45\textwidth}
  \centering
  \includegraphics[width=\linewidth]{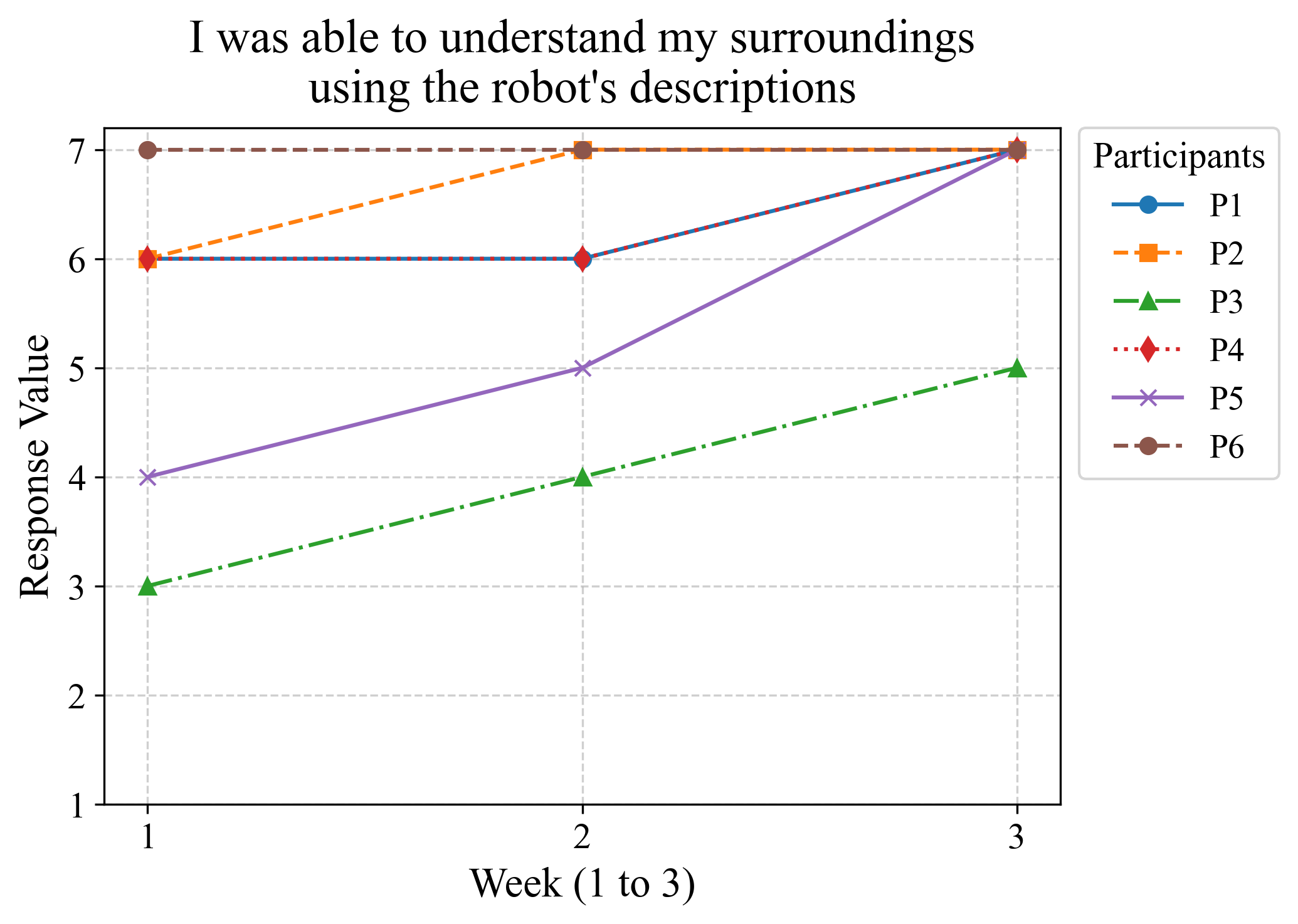}
  \caption{\textbf{Ratings for Statement (3): Ability to Understand Surroundings Using the Robot’s Descriptions.} Participant responses across three weeks showed consistently high or increasing agreement. P2 and P6 reported maximum scores early on, while P5 demonstrated the largest improvement over time.}
  \Description{Line chart showing participant responses across Weeks 1–3 to the statement “I was able to understand my surroundings using the robot’s descriptions.” P1 stays at 6 for Weeks 1–2, then increases to 7. P2 rises from 6 to 7 by Week 2 and remains at 7. P3 increases steadily from 3 to 5. P4 remains at 6 through Week 2, then rises to 7. P5 increases from 4 to 7. P6 remains at 7 across all weeks.
}
  \label{fig:surroundings_ratings}
\end{minipage}
\hfill
\begin{minipage}{0.45\textwidth}
  \centering
  \includegraphics[width=\linewidth]{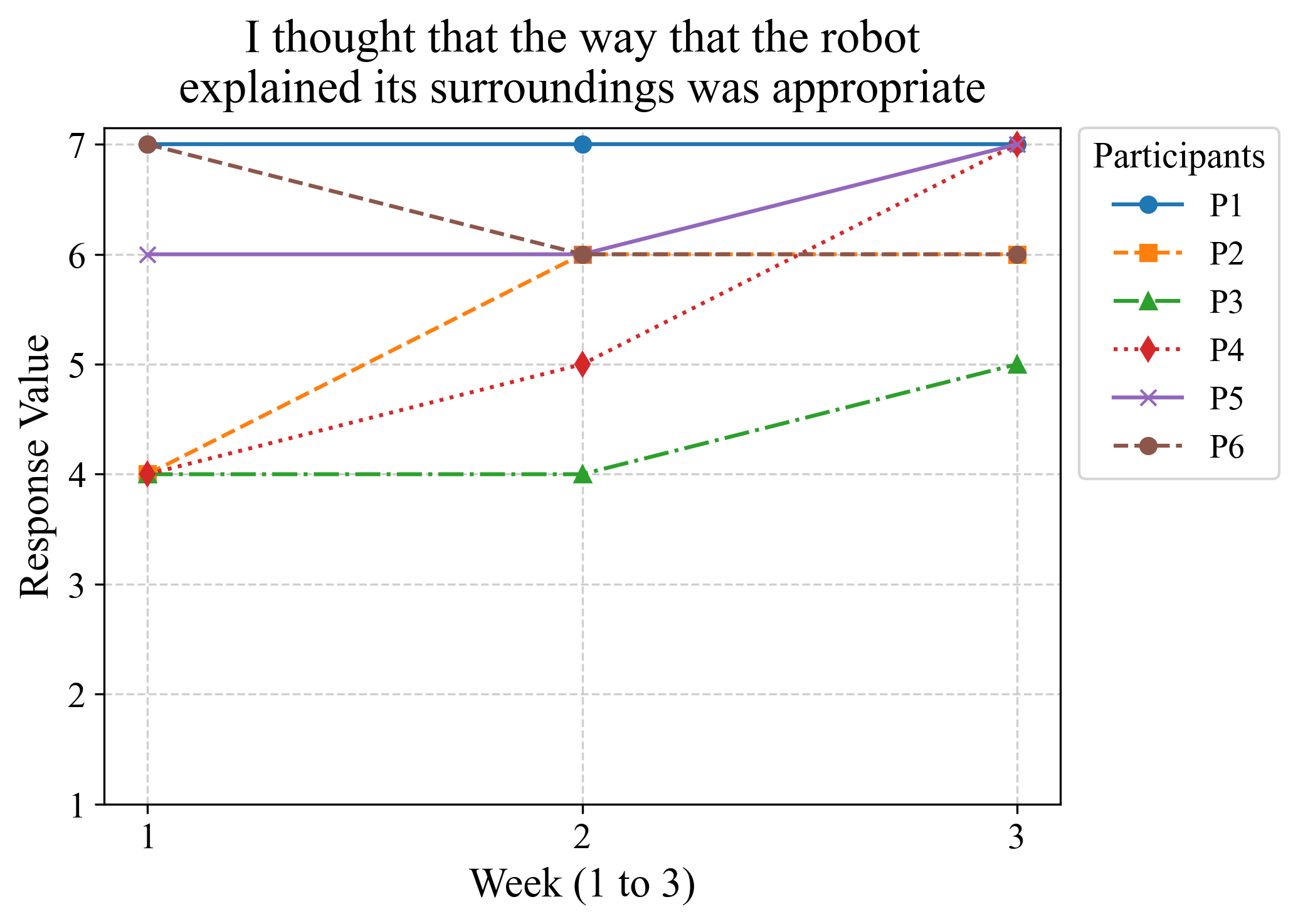}
  \caption{\textbf{Ratings for Statement (4): Appropriateness of the Robot’s Environmental Explanations.} Participant responses across three weeks were generally high, with P1 consistently rating the maximum, P4 showing the largest upward trend, and only P6 decreasing slightly from 7 to 6.}
  \Description{Line chart showing participant responses across Weeks 1–3 to the statement “I thought that the way that the robot explained its surroundings was appropriate.” P1 remains high at 7 throughout. P2 rises from 4 to 6 by Week 2 and stays constant. P3 stays at 4 through Week 2, then increases to 5. P4 increases from 4 to 7 across the three weeks. P5 rises steadily from 6 to 7. P6 decreases from 7 to 6 and remains stable.}
  \label{fig:explanations_ratings}
\end{minipage}
\end{figure*}

For statement (3), which asked whether participants were able to understand their surroundings using the robot’s descriptions, ratings also increased across the three weeks (Fig.~\ref{fig:surroundings_ratings}). Similarly, for statement (4), which asked whether the robot’s explanations were appropriate, participants expressed growing agreement (Fig.~\ref{fig:explanations_ratings}). \red{Linear mixed–effects analyses showed that Week was also a significant predictor for Statement~3 ($F(2,10) = 8.45$, $p = .0071$), indicating that participants increasingly felt able to understand their surroundings using the robot’s descriptions. For Statement~4, the effect of Week was positive but not significant ($F(2,10) = 2.50$, $p = .13$), showing a modest but less consistent increase in perceived appropriateness of the robot’s explanations.}

\begin{figure}[t]
  \centering
  \includegraphics[width=\columnwidth]{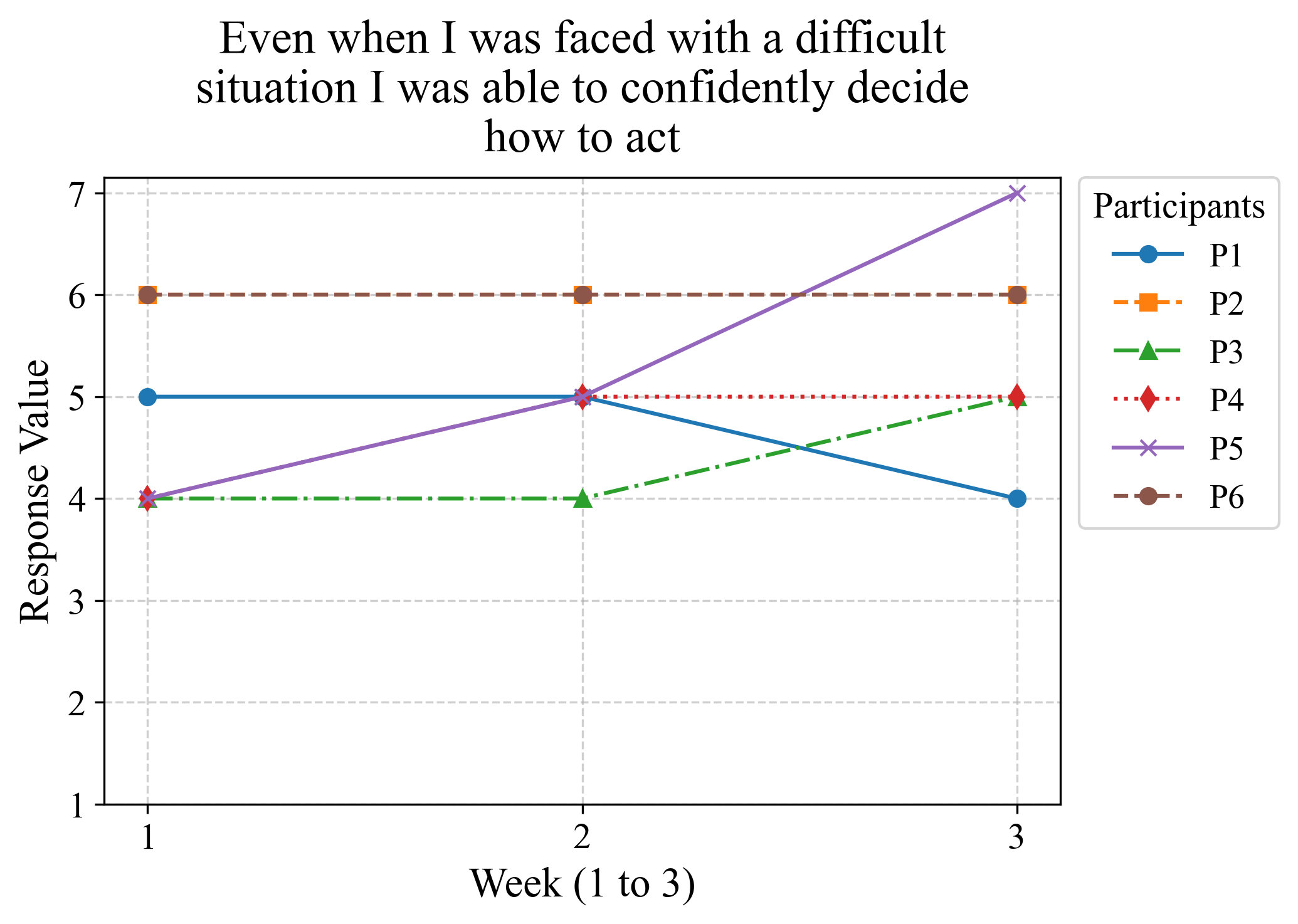}
  \caption{\textbf{Ratings for Statement (5): Participants’ Confidence in Acting While Using the Robot.} Responses over three weeks showed variation across individuals. P5 reported the strongest increase in confidence, P6 remained consistently high, while P1 declined slightly and others showed gradual improvements.}
  \Description{Line chart showing participant responses across Weeks 1–3 to the statement “Even when I was faced with a difficult situation I was able to confidently decide how to act.” P1 stays at 5 through Week 2 then decreases to 4. P2 stays constant at 6 across all weeks. P3 stays at 4 through Week 2 then increases to 5. P4 rises slightly from 4 to 5. P5 increases steadily from 4 to 7. P6 remains constant at 6 throughout.}
  \label{fig:confidence_ratings}
\end{figure}

We observed a trend in responses to statement (5), which asked whether participants felt able to act confidently while using the robot. Overall, most participants reported increased confidence over time, though there was some variability across individuals. P6 remained consistent across all three weeks, while P1 showed a decrease in confidence in Week 3. Notably, P5 shifted from being neutral in Week 1 to strongly agreeing by Week 3, indicating a substantial gain in confidence. Similarly, P4 moved from a neutral position in Week 1 to moderate agreement in Weeks 2 and 3. P2 and P3 both maintained stable confidence levels in Weeks 1 and 2, but trended upward in Week 3 (Fig.~\ref{fig:confidence_ratings}). \red{However, it is important to note that a mixed–effects model found that Week was not a significant predictor of confidence ($F(2,10) = 1.25$, $p = .33$). This indicates that although several participants showed descriptive increases in confidence over time, these changes were not consistent enough across the group to produce a reliable overall trend.}

% \begin{figure}[h]
%     \centering
%     \includegraphics[width=0.6\linewidth]{images/Even_when_I_was_faced_with_a_difficult_situation_I_was_able_to_confidently_decide_how_to_act.png}
%     \caption{Ratings for Statement (5): Participants’ Confidence in Acting While Using the Robot. Responses over three weeks showed variation across individuals. P5 reported the strongest increase in confidence, P6 remained consistently high, while P1 declined slightly and others showed gradual improvements.}
%     \label{fig:confidence_ratings}
% \end{figure}

In the answers to the RoSAS scale questions, participants were generally positive about the robot. On average, they showed high agreement with positive attributes such as being interactive, reliable, responsive, competent, and knowledgeable, and high disagreement with negative attributes such as scary, strange, awful, awkward, dangerous, and aggressive.  

\red{For the answers to the RoSAS scale,} there was no clear trend of consistent increase or decrease across weeks. Instead, some participants’ ratings fluctuated, going up in one week and down in another. Several of these changes were tied to specific circumstances participants encountered during that session, such as larger crowds, the presence of more foreign visitors, or the perception of younger children running around in the environment. 

% \red{While users could be given ways to amplify their own voices, several participants explained that they preferred not to initiate social interactions themselves (P4). For example, P4 described the robot as a way to manage these situations without having to speak directly. Providing the option to delegate therefore supported both users who preferred independence and those who preferred the robot to take the lead.}

\subsection{Dynamic Crowds and Real-World Occurrences}

In addition to the obstacles and crowds staged by actors, we observed several natural interactions between participants and other museum visitors. In some cases, people noticed the robot approaching and moved aside on their own. When visitors had their backs turned and did not see the robot, participants often relied on the robot’s voice to request that people move, which typically prompted a response.  

There were also instances where visitors did not understand the robot’s request. Several participants noted that the museum had a large number of foreign visitors who were not fluent in the local language, raising concerns that the robot’s verbal prompts might go unnoticed or unrecognized. Participants suggested that the system could be improved by offering the ability to switch languages or by using additional non-verbal signals, such as auditory cues (e.g., beeps), to attract attention and ensure that surrounding people understood the robot’s presence and intention.

\subsection{Unanticipated Obstacles and User Responses}  

Several participants experienced unexpected behavior from the robot due to limitations in its path planning. In multiple instances during Week 3 on Route 3, the robot struggled to navigate around a slanted wall that angled toward the walkway. The robot’s sensors did not detect the wall until it was too close, causing it to become completely stuck. In these cases, the GPT-based obstacle detection repeatedly announced: “There is an obstacle. There is a wall.”

After calling for help, either via the robot or by using their own voice, participants then had to verbally indicate that they were stuck and request assistance. Actors approached and manually moved the robot only after participants explained the situation. During the post-study interviews, several participants reflected that if this occurred in a real-world setting, they would feel uncomfortable asking strangers for help, especially given the robot’s weight. While they acknowledged the difficulty of such situations, participants were uncertain how they might manage them outside the study environment.

\subsection{Continued Usage}

An important aspect of our longitudinal study was understanding how participants’ perceptions and comfort with the robot changed over multiple sessions. Experiencing the system across several days gave participants a clearer sense of how it might fit into their everyday routines. During the post-study interviews in the final week, we asked whether they would want to continue using the robot and in what contexts. 

Participants offered varied perspectives. Several noted that they would not use the robot constantly but preferred it for situations that are more challenging or cognitively demanding. For example, P1 reported that the robot would be helpful in “crowded or complex places, like a mall,” where locating items and moving safely can be difficult. Others mentioned supermarkets, where dynamic obstacles and tight layouts often present challenges. One participant (P6) suggested the robot would be useful in environments that require careful navigation, such as “somewhere like a museum where you need to be careful about the objects around you.” Others described interest in using the robot in outdoor markets or small shop areas, where paths are irregular and human assistance may be inconsistent.

While participants did not view the robot as an all-purpose daily companion, they recognized clear situational value. Because the study exposed them to a variety of social navigation scenarios, participants were able to consider how the robot could support different navigation demands in their daily lives. Overall, participants expressed interest in using the robot selectively in more demanding contexts, highlighting its potential role as a \textit{contextual support tool} rather than a constant assistant.

\section{Discussion}

\subsection{Delegating Social Interaction to the Robot}

The results demonstrate that participants’ delegation to the robot was shaped both by individual differences and situational demands, as evidenced by many other similar systems~\cite{wei2025human}. Participants who began with high levels of delegation (P2, P3, P4) maintained this reliance across all three weeks, suggesting that for some, the robot immediately served as a valued partner in managing social interactions. Others, such as P1, P5, and P6, initially resisted delegating to the robot, emphasizing independence or relying on long-standing habits of asking for help themselves. However, over time these participants exhibited tendencies to delegate to the robot, particularly in noisy or crowded contexts where their own voices were less effective. \red{While users could be given ways to amplify their own voices, several participants explained that they preferred not to initiate social interactions themselves. For example, P4 described the robot as a way to manage the situation without having to speak directly. Providing the option to delegate therefore supported both users who wanted to handle interactions independently and those who preferred the robot to take the lead.} Prior work has noted that blind individuals often negotiate boundaries to avoid drawing unwanted attention from those around them~\cite{worth2013visual}. In contrast, our findings show that the robot provided a way for participants to call attention to themselves in order to accomplish a task, thereby allowing them to sidestep the discomfort of initiating a direct social interaction. This progression illustrates a process of trust-building and pragmatic adaptation, in which participants came to view the robot as a reliable partner for managing social interactions.

Importantly, participants emphasized that the robot’s synthetic voice, though not natural, was effective because it projected loudly and captured attention. Prior work similarly demonstrates that synthetic or machine-generated voices, when well-designed, can match or even outperform human voices in noisy environments by improving clarity, intelligibility, and attention capture~\cite{hennig2012expressive}. 

However, some participants raised concerns about phrasing and comprehension in multilingual or international settings, noting that visitors who did not understand the language sometimes failed to respond. These observations highlight the need for adaptable voice styles and language flexibility in future systems.

\red{These findings show why delegation must be examined in longitudinal settings rather than single-session studies. Initial behaviors did not always reflect stable preferences; some participants who began with strong independence later adopted selective delegation as they gained experience, while delegation-first participants remained consistent across weeks. Longitudinal study designs reveal how preferences stabilize or shift as users build confidence in the system and understand its capabilities in real-world conditions.}

Across participants, we observed three broad personality types shaping user delegation behavior:  

\begin{enumerate}
    \item \textbf{Independent-first participants.} Individuals such as P1 and P6 described themselves as wanting to maintain independence and “do everything that sighted people can do.” They were initially reluctant to delegate to the robot, preferring to take initiative themselves. However, with longer exposure they recognized scenarios where their own voices were less effective, such as  in noisy or crowded spaces, and began to selectively use the robot.  

    \item \textbf{Balanced participants.} A second group showed a more flexible approach from the outset, combining independent action with willingness to delegate to the robot when appropriate. These participants expressed comfort switching between modes from the start; however, they tended to use the robot more at the beginning.   

    \item \textbf{Delegation-first participants.} Participants like P4, who described themselves as shy or uncomfortable with asking strangers for help, relied on the robot consistently. For them, the robot lowered the social barrier to requesting assistance, enabling interactions they might otherwise have avoided.  
\end{enumerate}

These personas show that user delegation to the robot is both situational and personality-dependent: while some participants required time and repeated exposure before embracing the robot, others valued its role immediately as a social intermediary. The diversity of strategies highlights the importance of designing navigation-assistive robots that can flexibly support different interaction styles and evolve with users over time.

A prior study that examined user preferences for receiving environmental information highlighted the need to investigate how shared control between humans and robots might function when information is provided by an automated system rather than by a human in a Wizard-of-Oz (WoZ) setting~\cite{kamikubo2025beyond}. In our study, we found that when participants queried general information about their surroundings using GPT-generated descriptions, they typically viewed this feature as most useful for gaining a broad sense of the environment or the overall ``vibe,'' rather than for highly localized details about what was directly in front of them. At the same time, we observed that some participants, particularly P6, began to repurpose the "Surrounding GPT" feature over time, increasingly activating it after being alerted to an obstacle. This shift suggests that the function not only supported orientation but also became a strategic tool for interpreting the robot’s behavior and deciding when to intervene.

\red{Across the study, early behaviors did not always match later preferences. P1, who initially rejected delegation, later found the robot more effective at asking others to move, while P4 consistently preferred the robot to take the lead. Balanced participants remained flexible throughout. Despite these differences, all participants eventually expressed confidence in the robot’s ability to handle social requests, showing that repeated exposure strengthens trust and reliance beyond initial personality-based tendencies.}

\subsection{Learning and Adaptation Over Time}

Prior work has highlighted the importance of examining long-term use when investigating how blind users interact with navigation-assistive robots, noting that preferences and strategies may evolve as familiarity increases~\cite{kamikubo2025beyond}. Our findings build on this by showing that user adaptation was not simply a byproduct of repeated exposure, but instead reflected an evolving process of learning when and how to collaborate effectively with the robot. Our results extend prior work by showing how longitudinal exposure shapes users’ understanding of the robot and alters their viewpoints over time. Repeated sessions revealed that participants often revised their initial opinions as they gained experience, demonstrating that preferences observed in one-time encounters may not reflect stable judgments.

Participants shifted from exploratory use of the robot’s features to more deliberate, context-sensitive strategies. For example, P1 initially rejected delegation to the robot altogether, framing independence as a matter of personal identity, yet later employed the robot selectively in noisy settings (Fig.~\ref{fig:help_ratings} and Fig.~\ref{fig:move_ratings}). Similarly, P6 moved from no delegation to the robot in Week~1 to steadily greater reliance by Week~3 (Table~\ref{tab:robot_social_use}), while P5 transitioned from primarily using his own voice to relying almost entirely on the robot. These trajectories illustrate how experiential learning, combined with situational pressures such as crowd density and background noise, fostered gradual trust in the robot’s social capabilities and shifted the boundaries of when users considered it advantageous to delegate social interactions to the robot.  

Participants also reported developing interpretive frameworks for the robot’s behavior, such as recognizing corrective movements (e.g., backing up, pivoting, repeating small motions) as signs that the robot was “searching” for a path rather than fully stuck. This interpretability directly shaped collaboration: rather than intervening immediately, some participants chose to wait until the robot appeared unable to progress, suggesting that trust was dynamically calibrated in response to perceived robot effort (Table~\ref{tab:obstacle_action_compressed}). In parallel, use of the ``Surrounding GPT'' function evolved from curiosity-driven exploration in early sessions to targeted requests for context-specific information in later sessions (Table~\ref{tab:surrounding_gpt}). This reflects a broader process of experiential learning in which participants refined both their delegation strategies and their sense of when environmental descriptions provided actionable support.  

Confidence ratings add further nuance to these findings (Fig.~\ref{fig:confidence_ratings}). While participants such as P5 and P4 reported increased confidence over time, attributing this to growing familiarity with the robot’s movements and descriptions, others (P1, P6) remained stable or showed declines in Week~3. These fluctuations underscore the influence of situational variability: differences in crowd density, background noise, and the unpredictability of museum environments all shaped how confident participants felt in acting alongside the robot. Importantly, confidence appeared to emerge from a combination of two processes: developing interpretive understanding of the robot’s signals and receiving timely, interpretable information about the surrounding environment.  

Taken together, these findings suggest that collaboration with navigation-assistive robots is a dynamic, context-sensitive process shaped by both user learning and environmental contingencies. Adaptation occurred not only through repeated interaction but also through participants’ active construction of mental models about the robot’s behavior. Designing systems that explicitly support this interpretive process—by making corrective movements more transparent, providing adjustable levels of descriptive detail, and scaffolding confidence across longer-term use—may strengthen trust, sustain user agency, and ultimately promote real-world adoption.

\subsection{Limitations and Implications}

Our findings show that collaboration between user and robot is a dynamic process shaped by trust, learning, and situational demands. This underscores the importance of long-term, real-world evaluation: participants needed time not only to learn how the system functioned but also to refine their own preferences. Unplanned events in the museum, such as language mismatches, unexpected obstacles, and mapping errors, created challenges unlikely to appear in controlled settings.  

\red{ In our system, the robot attempted to reroute when possible, but many scenarios intentionally focused on moments where rerouting was not feasible, reflecting the freezing robot problem and the need for user decision-making during social navigation. However,even if future systems can autonomously plan alternative paths around obstacles or crowded areas ~\cite{liu2021tactile,xu2020virtual}, and not face the freezing robot problem, such rerouting may result in longer travel times or paths that do not align with a user’s preferences. Users may still want control over where they go and how they navigate. Future work could examine whether, in longitudinal settings, participants would choose to rely on the robot to move a crowd rather than accept a longer detour. Our findings identified several scenarios in which blind participants preferred having the robot ask people to move, and showed how this preference varied across personality types and situational demands. Building on this, future systems could offer both options in real time, allowing users to compare the cost of a detour with the effort of engaging socially through the robot.}

% Several key limitations emerged. First, the robot often struggled to distinguish between lines and crowds, leaving participants uncertain about how to act. Prior work shows that misinterpreting social context undermines trust~\cite{clark2019makes}, a concern echoed by participants who wanted clearer communication of obstacles. Second, the lack of interactive question-and-answer functionality limited environmental descriptions. While participants appreciated GPT-generated summaries, they wanted the ability to query specific details, aligning with research on user-driven explanations that better support decision-making~\cite{yu2023perception,hernandez2023explaining}. Third, unplanned obstacles sometimes left the robot stuck (e.g., against a slanted wall), with participants noting that moving a heavy robot would be impractical without robust recovery mechanisms such as rerouting or safe assistance requests. Finally, language barriers reduced effectiveness in multilingual contexts, though participants still valued the robot’s louder, more authoritative voice over their own, even when comprehension was uncertain~\cite{kayukawa2019bbeep}.  

\red{Several limitations emerged. First, the robot often struggled to distinguish between lines and crowds, leaving participants unsure whether they should wait or initiate an interaction. Prior work similarly notes that identifying socially meaningful group formations from a single image using off-the-shelf LLMs or vision–language models remains a challenging open problem in accessibility and computer vision, particularly when subtle motion cues are needed to interpret intent~\cite{zhang2025escalator}. These difficulties help explain why the system could not consistently communicate such distinctions to participants.}

\red{Second, although participants appreciated GPT-generated summaries, the system did not support interactive question-and-answer functionality. We intentionally did not implement Q\&A because LLM inference latency, combined with the hardware constraints of an on-device system and inconsistent network speeds needed for online processing, would have caused noticeable delays during navigation. However, interactive querying remains a promising direction for future work as models become faster, lighter, and better suited to real-time deployment on embedded platforms.}

Third, unplanned obstacles occasionally caused the robot to become physically stuck (for example, against slanted walls or tight corner geometries). Participants noted that manually repositioning a heavy robot would be impractical, highlighting the need for more robust recovery mechanisms such as autonomous rerouting or more reliable assistance prompts.

Finally, language barriers sometimes reduced effectiveness in multilingual environments. Even so, participants reported that the robot’s louder and more noticeable voice often helped draw attention better than their own speech, even when bystander comprehension was not guaranteed~\cite{kayukawa2019bbeep}.

These limitations point to clear design directions. Navigation-assistive robots should (1) make obstacle states and corrective actions more transparent, (2) support user-driven queries for adaptive levels of detail, (3) provide flexible, multilingual communication, and (4) incorporate recovery strategies for unresolvable obstacles.  

\section{Conclusion}

This work examined how blind participants engaged with a navigation-assistive robot over three weeks, focusing on how delegation and collaboration evolved through repeated interaction. Participants shifted from exploratory use to more deliberate strategies, selectively delegating social interactions to the robot and using GPT-generated descriptions to guide decision-making. These behaviors were shaped by both individual preferences and situational demands such as noise, crowd density, and the robot’s corrective movements.  

Our findings show that delegation is dynamic and adaptive: users develop trust, form mental models of the robot’s behavior, and refine strategies over time. \textbf{While some users may prefer to handle social interactions independently, others may consistently rely on the robot, or adapt their approach depending on the circumstances.} Allowing for this flexibility is critical, as preferences vary widely across individuals and contexts. \red{Therefore, our study demonstrates that assistive navigation systems for blind individuals should be evaluated in longitudinal settings, since preferences and usage patterns evolve over time and may differ substantially from those observed in first-time encounters.}

% \textbf{Therefore, ourLongitudinal evaluation is therefore essential, as trust, confidence, and reliance on robot support evolve only through repeated use in real-world conditions.}  

Future work should build adaptive systems that personalize support, for example by enabling Q\&A-style environmental descriptions, context-sensitive speech, or models that learn from prior user behavior. Extending this research to longer deployments and broader contexts such as outdoor navigation or public transit will further test how these dynamics scale. \red{Additionally, equipping models that are more accurate at detecting obstacles may influence delegation behavior and should be further investigated. }

Taken together, these findings show that the design of navigation-assistive robots must go beyond technical autonomy alone. Instead, robots must be built as adaptive partners that negotiate control fluidly with their users, supporting independence while remaining responsive to context. By prioritizing flexibility, transparency, and personalization, future systems can more effectively empower blind individuals and promote confident, long-term use.  
\bibliographystyle{ACM-Reference-Format}
\bibliography{cleaned_reference_full}

\appendix

\end{document}